# Graphene-based integrated photonics for next-generation datacom and telecom

Marco Romagnoli[1], Vito Sorianello[1], Michele Midrio[2], Frank H. L. Koppens[3,4], Cedric Huyghebaert[5], Daniel Neumaier[6], Paola Galli[7], Wolfgang Templ[8], Antonio D'Errico[9] and Andrea C. Ferrari[10]

[1]CNIT, Photonics Networks and technologies Laboratory, Pisa, Italy.
[2]CNIT, University of Udine, Udine, Italy.
[3]ICFO, Institut de Ciencies Fotoniques, the Barcelona Institute of Science and tTechnology, Castelldefels, Spain.
[4]ICREA, Institució Catalana de Recerça i Wstudis Avancats, Barcelona, Spain.
[5]IMEC, Leuven, Belgium.
[6]Advanced Microelectronic Center Aachen, AMO GmbH, Aachen, Germany.
[7]Nokia italia, vimercate, italy.
[8]Nokia Deutschland AG, Bell Laboratories, Stuttgart, Germany.
[9]Ericsson Research, Pisa, Italy.
[10]Cambridge Graphene Centre, Cambridge University, Cambridge, UK.

**Abstract**

Graphene is an ideal material for optoelectronic applications. Its photonic properties give several advantages and complementarities over Si photonics. For example, graphene enables both electro-absorption and electro-refraction modulation with an electro-optical index change exceeding $10^{-3}$. It can be used for optical add–drop multiplexing with voltage control, eliminating the current dissipation used for the thermal detuning of microresonators, and for thermoelectric-based ultrafast optical detectors that generate a voltage without transimpedance amplifiers. Here, we present our vision for graphene-based integrated photonics. We review graphene-based transceivers and compare them with existing technologies. Strategies for improving power consumption, manufacturability and wafer-scale integration are addressed. We outline a roadmap of the technological requirements to meet the demands of the datacom and telecom markets. We show that graphene based integrated photonics could enable ultrahigh spatial bandwidth density , low power consumption for board connectivity and connectivity between data centres, access networks and metropolitan, core, regional and long- haul optical communications.

**Introduction**

In the past 25 years, data traffic has exponentially grown, and optical fibre amplifiers have greatly contributed to this expansion in traffic[1]. Optical fibre amplifiers enabled the Internet era, offering the faster data rates required for smart phones and social media[2]. The next wireless communication technology, known as 5G (fifth generation)[3], requires an increase in bandwidth of three orders of magnitude (>500 Mb s−1) for each user and all objects connected to the Internet (ref.[4]), as the 5G evolution is driven by the growing mobile communication markets and the development of the Internet of Things (IoT)[5]. This growth in communication is predicted to increase the global gross domestic product to ~US$1.9 trillion[6], with ~50 billion connected devices in use by 2020 (ref.[7]). Therefore, there is urgent demand for a technology that can meet requirements in terms of bandwidth and power consumption. Considering these growth projections, one needs to be mindful of the impact of communication technologies on global energy consumption and global warming[8]. At present, the information and communication technology (ICT) industry accounts for 2–2.5% of all greenhouse emissions, according to the International Telecommunications Union, and this is predicted to increase to ~4% by 2023.

Photonics is poised to play an increasingly important role in ICT (Fig. 1a), since the fixed high capacity links are largely based on photonic technologies. Photonic devices need to support ultra-large bandwidth operation, for example, 200 Tb s−1 in a single fibre[9] and >10 Tb $s^{-1}$ $cm^{-2}$ in integrated Si photonics chips[10]. To achieve this, the key components of Si photonics, photodetectors and modulators, need very high performances in terms of speed (≥25 Gb $s^{-1}$), footprint (<1 $mm^2$), insertion loss (<4 dB), manufacturability (>$10^6$ pieces per year) and power consumption (<1 pJ $bit^{-1}$). To date, these requirements have not been fulfilled in one system[11]. Furthermore, in terms of production volumes, photonics is not yet comparable to microelectronics[12], even if the increase in demand for optical networks would, by 2021, lead to an average global Internet Protocol (IP) traffic of 3.3 ZB (zettabytes), corresponding to an average usage data rate of ~800 Tb $s^{-1}$ (refs[13,14]). In the context of the IoT[7], other applications, including infrared (IR) sensors, biosensors, environmental sensors, metrology, quantum communications and machine vision, will require even larger production volumes[13,15].

The telecom network can be divided into three segments: access, aggregation and core (Fig. 1a). The access network is the interface between subscribers and the immediate service provider. The aggregation network aggregates all the input data streams from tributary access networks, converging towards the higher-level core network. The aggregation network includes local and metropolitan networks, which then converge to regional networks. A local area network (LAN) interconnects computers within a limited area, such as a residence, school, laboratory, university or office building. A metropolitan area network (MAN) interconnects users with computer resources or communication servers in a geographical area or region larger than that covered by even a large LAN. A regional network covers areas of radii from approximately 10 km to 500 km. The core network is the set of communication facilities that interconnect primary nodes, delivering routes to exchange information between various sub-networks.

The points of access to the access network are the land-line and wireless individual subscriber networks and the radio base stations for wireless communications. Signal routing occurs in data centres located in all segments of the communication network. In total, maintenance and evolution of the telecom network requires >1 million devices per year[16,17]. All equipment supporting very high bandwidths, such as the 100GE (Gigabit Ethernet)[18,19] electro-optical interfaces (that is, converters from electrical to optical signals or vice versa), and with high connection capacity for access networks, aggregation networks and data centre interconnections is based on optical technologies (Appendix 1). The global optical high-capacity transceivers market is estimated to reach ~$6.87 billion by 2022, driven by the availability and cost-effectiveness of devices with speeds between 100 and 400 Gb s$^{-1}$ (ref.[20]).

Data traffic at the periphery of the communication network originates from devices with IP addresses (for example, laptops, surveillance cameras and smart phones) (Fig. 1b), and it is expected to increase at a rate of ~1.6 billion connected devices per year — projected to be ~12.5 billion by 2020. Photonic technologies are increasingly playing a role in access networks. Applications range from fibre to the home (FTTH) scenarios to the backhauling of wireless nodes (e.g. access nodes or base stations)[21].

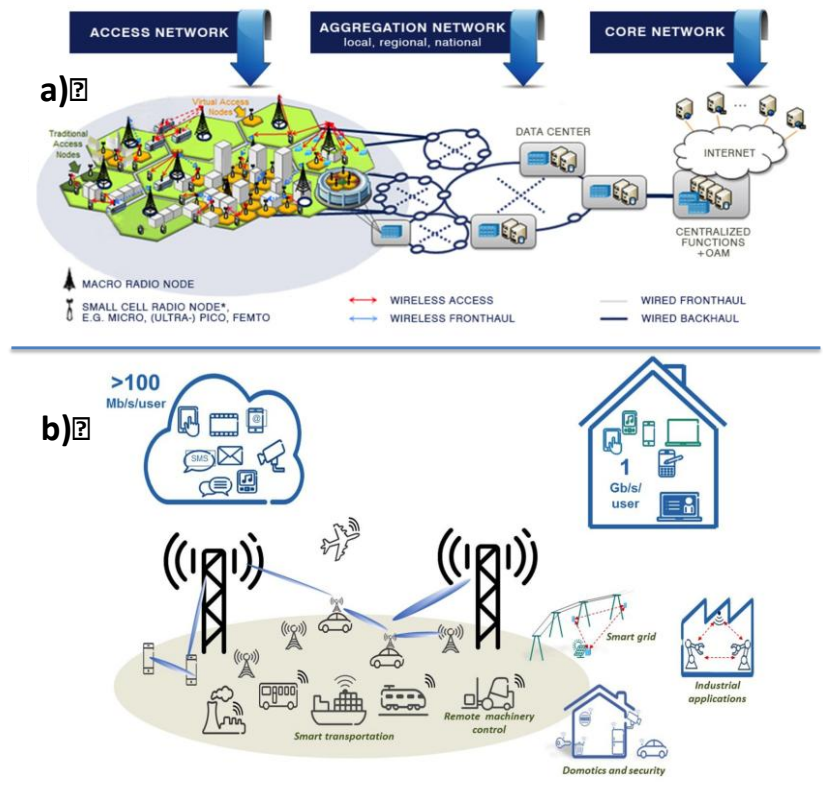

**Fig. 1.** | **The evolution of communications**. **a** | Widespread communication scenario indicating how the telecom network can be divided into three parts: access, aggregation and core. Users can connect to the telecom network via the access network and ask for a service delivered by other parts of the network, where data centres are connected to perform dedicated applications. The data centres connected to the core network have higher computational resources than those connected to the aggregation network or those directly connected to the access network, such as a stadium communication network. **b** | Schematic depiction of the fifth-generation wireless system, 5G. To enable 5G, all the available network infrastructures must evolve with new levels of flexibility and automation (more specifically, networks performing self-operations, optimization and healing), with higher priority given to network optimization, security, energy and cost efficiency. A large number of different objects with IP addresses will be controlled, monitored and connected through the 5G network. A description of the evolution of mobile communication networks with the advent of the 5G era can be found in Appendix 5.

The emergence of IoT and, eventually, the Internet of Everything (IoE) requires intelligent management of the huge network of interconnected 'things', according to the International Telecommunications Union. The IoT vision is for ubiquitous 'smart objects' to exchange information anywhere and anytime using their individual IP addresses. 'Smart' is defined as sensing combined with decision-making and artificial intelligence[22]. Over the past 15 years, the price of sensors, processors and networking has decreased according to Moore's Law, giving rise to new products arising from interconnected machines and devices (or 'things') via the network. The consequences of such technologies are far beyond the individual cases and have the potential to change our society, as the internet has done. Thanks to the widespread deployment of WiFi, it is easy to add new networked devices to the home, office or other locations. The adoption of IPv6 enables an almost unlimited number of devices to be connected to networks[23]. Major system vendors and market forecasts estimate ~28 billion connected devices by 2021 (ref.[21]) and ~100 billion by 2025 (ref.[24]), with a market of up to tens of trillions of dollars by 2025 (ref.[25]), leading to a 'smart' societal change.

The increase in the number of connected devices requires a large and pervasive photonic communication infrastructure (Fig. 1a), with an optical bandwidth considerably greater than 25 THz in installed optical fibre systems[26,27]. The 5G network will need energy-efficient cells (10–100 times more energy efficient than 4G), based on photonic[28] or millimetre wave connection for fronthaul and backhaul transport[29]. Fronthaul transport is implemented between the centralized baseband units and the relevant remote radio units to enable a seamless connection without affecting radio performance. Backhaul transport realizes connections from the centralized baseband units to the IP core network to perform end- to-end solutions for low latency (that is, the time delay due to processing)[30]. Also under consideration is millimetre- wave-based ultrahigh capacity (>1 Gb $s^{-1}$ per user and >10 Gb $s^{-1}$ per remote antenna unit), short-range (<1 km) access[31 . As a result, there is a need for photonic interconnections that are cost efficient (<\$10/Gb $s^{-1}$ by 2020 and <\$1/Gb $s^{-1}$ by 2025)[32,33] and have large channel bandwidths (>100 GHz). A description of the evolution of mobile communication networks with the advent of the 5G era can be found in Appendix 5.

At present, optical interconnections in data centres are mainly between boards that provide the platform on which the electronic components and optical or electrooptical devices are connected. In the near future, the number of optical interconnections will increase[34]. As a result, by 2021, the production of optical interconnections is predicted to be >10 million per year[35]. The photonic devices — most commonly, modulators (Appendix 1; Appendix 2) and detectors (Appendix 2) — needed to meet these requirements are based on $LiNbO_3$ (refs[36,37]), semiconductors such as InGaAsP/InP (refs[38,39]) and those used in Si photonics[11] (Table 1). Devices based on $LiNbO_3$ and semiconductors are established[40], whereas Si photonics is a newer and faster-growing field[17,32]. The parameters used to compare modulators are modulation efficiency, insertion loss and the figure of merit (FOM) for a phase shifting functionality ($FOM_{PM}$) (Appendix 2). The modulation efficiency of interferometer-based modulators is defined as $V_{\pi}L$, where $V_{\pi}$ is the voltage required to achieve a π phase shift of the optical carrier and L is the length of the phase shifter.

| Material | $V_{\pi}L$ (V mm) | Insertion loss (dB $mm^{-1}$) | $FOM_{PM}$ (V dB) | Refs |
|---|---|---|---|---|
| LiNbO3 (E) | 50–100 | 0.4 | 20–40 | 36 |
| LiNbO3 (E)[a] | 18 | 0.3 | 5.4 | 37 |
| InGaAsP/InP (E) | 5–10 | 0.7 | 3.5–7 | 38,39 |
| Si photonics (E) | 10–20 | 1–2 | 10–20 | 94–103,125 |
| Graphene (T) | 0.7–2.8 | 0.1–1.2 | 1–2 | 72,79,87 |

**Table 1** | Comparison of modulators based on different material platforms. E, experiment; T, theory; a Small- mode $LiNbO_3$ rib waveguide (width = 900 nm, rib height = 400 nm and slab thickness = 300 nm).

$LiNbO_3$ Mach–Zehnder modulators have low (<0.4 dB mm) insertion loss and high (>50 V mm) $V_{\pi}L$. According to Table 1 the device length needed to achieve a π phase shift with a driving voltage of 2 V is between 2.5 and 5 cm. InGaAs/InP detectors (ref.[40]) or Ge/Si (ref.[41]) have comparable performances in terms of both responsivity $R_{ph}$ (~1 A $W^{-1}$) and bandwidth (>40 GHz) and, at present, have a higher $R_{ph}$ than single-layer graphene (SLG) photothermal detectors, which have so far demonstrated[42] $R_{ph}$ ~0.36 A $W^{-1}$, and up to 100 GHz bandwidth[43]. However, for technologies to become widespread, devices must be mass produced, cost efficient, reproducible, reliable, and compliant with existing semiconductor processes and environmental regulations. With these considerations in mind, for large- scale production, Si photonic[11] devices are preferable to InGaAsP/InP ones because the technological processes are the same as those already present in Si foundries commonly used in the semiconductor industry. Thus, the Si photonics platform for single- wavelength components is a practical technology and permits co- packing of electronic functions with light sources[44]. Given that graphene photonics is compatible with Si photonics and other materials such as SiN and $SiO_2$, in the following we focus on the potential for integration of graphene with Si-based technology.

A high-performance photonic device requires highprecision fabrication equipment. For example, an optical lithography node size of 65 nm in a 300 mm fab (a semiconductor fabrication plant) provides a good trade-off between performance and cost, even if optochip costs are only ~20% of today's transceiver costs[44]. For Si photonics, Si- on-insulator (SOI) is used, costing >\$1,000 (at 2017 prices), five times more than Si (<\$200 at 2017 prices). Considering the ethernet roadmap to 2020, the important components for Si photonics — photodetectors and modulators — must be high performing in terms of speed (>50 Gb $s^{-1}$), footprint (<100 $\mu m^2$), insertion loss (≤1 dB) and energy consumption (~100 $\mu W$ $GHz^{-1}$). To date, these parameters cannot be satisfied in one system because of the trade-off between electro-optical properties and loss[46].

In Si photonics, light with wavelengths between 1,300 and 1,550 nm is guided by Si and is detected by Ge p–i–n photodetectors integrated on Si (ref.[41]). Graphene exhibits both electro-absorption[47] and electro-refraction[48] and, hence, can be used for light modulation and photodetection[49]. There remains a need for optical transmitter and receiver modules (transceivers) integrating waveguides with photodetectors and modulators on one chip, in parallel with wafer- scale processing[50].

The potential of graphene for photonics and optoelectronics has been discussed in previous reviews[49,51–54]. Here, we focus on the key arguments underpinning the development of graphene-based integrated photonics for high-speed datacom and telecom. Graphene-based photodetectors will remove the need for Ge epitaxy, which is currently used for Si photonics photodetectors, by replacing the Ge p–i–n photodetectors with a single or double SLG. Graphene photodetectors are not spectrally limited[55], unlike Ge p–i–n photodetectors[41], which operate below wavelengths of 1,600 nm. Graphene-based photodetectors can reach bandwidths of ~260 GHz (ref.[56]), as a consequence of the high carrier mobility, μ, of SLG (Appendix 3). Moreover, in voltage-detection mode, graphene-based photodetectors can function at zero dark current[57].

Another key component of transceivers is the electrooptic modulator. Graphene-based modulators have advantages over Si-based modulators. They are capable of broadband 30 GHz electro- absorption operation[58] based on modulating the resonance of a micro-ring resonator in and out of the critical coupling condition, they are compatible with complementary metal oxide semiconductor (CMOS) processing, and enable postprocessing fabrication and the use of different substrates. SLG does not require Si or Ge doping. Hence, the waveguide can be Si, SiN, $SiO_2$ or another transparent material. Practically, this implies a post-processing shift in manufacturing from front-end to back-end-of-line. In addition, graphene technology does not necessarily require expensive SOI wafers, or implantation for junctions, and Ge growth for detectors. Because SiN and $SiO_2$ waveguides are wider than Si photonics ones, the lithography node can be relaxed. The waveguide size is ~0.5 μm for Si, ~1.5 μm for SiN and ~ 8μm for $SiO_2$. All these factors will simplify the technology and reduce costs, making small and medium production volumes more affordable because the initial non- recurring engineering is less than in a SOI- based line. This means that the volume threshold to implement a product in a Si fab can be reduced, enabling the cost- effectiveness of mediumvolume products (10,000–100,000 chips per year), thus opening up markets wider than those for consumer electronic products, which require higher volumes.

**Graphene-based modulators**

The basic components of communication systems are waveguides, modulators and photodetectors. Modulation of light is one of the key operations in photonic integrated circuits[59] (Appendix 1). The properties of guided light, such as amplitude, phase and polarization, can be modulated by altering specific properties of the guiding medium. For example, electro-absorption modulators[59,60] induce modulation of the amplitude of the propagating light through the modulation of the optical absorption of the waveguide. Electro-refractive modulators[61,62] alter the phase of the propagating light by changing the effective index neff (the index of refraction determining the phase velocity of light in a waveguide)[63]. The Franz–Keldysh effect[60,64–66] (Appendix 4) and the quantum-confined Stark effect[67,68] (Appendix 4) can be exploited in electro-absorption modulators.

SLG is a broadband absorber with 2.3% absorption at any wavelength at normal incidence, as a consequence of its lack of a bandgap[69]. This corresponds to $10\log_{10}(1-0.023)$ ~0.1 dB. By superimposing SLG on a Si waveguide, it is possible to enhance the interaction of SLG with light[70]. Absorption increases significantly from ~0.1 dB at normal incidence to thousands of dB cm$^{-1}$ when light propagates along the waveguide[71,72]. If the Fermi energy, $E_F$, of SLG is shifted, the absorption is reduced or cancelled[73]. When the $E_F$ is larger than the energy of the propagating photons, $E_{ph}$, SLG is more transparent as a result of Pauli blocking[74] and the index of refraction changes[70]. The electrical and optical properties of SLG depend on carrier concentration. Defects can give a background loss independent of $E_F$ and introduce losses[70], thus degrading the electro-optical properties. In the Kubo model[75], the presence of defects can be taken into account by introducing intraband transitions due to long- range scattering, owing to the presence of impurities, trap states and screening[76], accounted for by the scattering time, τ[76] (Appendix 3).

To illustrate the effect of τ on background loss, we consider single SLG and double SLG devices operating at 1,550 nm (corresponding to $E_{ph} \approx 0.8$ eV) (Fig. 2). For the single SLG device, one layer of SLG is placed on a Si-ridge waveguide with core dimensions of ~480 nm (width) x 220 nm (height) on top of a 60 nm-thick Si slab (Fig. 2a). A 5 nm-thick dielectric layer is placed between Si and the SLG. The slab waveguide and SLG have electrical contacts on their surface, and the result is a Si–insulator–SLG capacitor[70,71]. In the double SLG device, two SLGs are placed on an undoped Si waveguide (Fig. 2c). One of the SLGs is separated from the Si waveguide by a 5 nm-thick dielectric layer. Above this SLG is an additional 5 nm-thick dielectric layer, followed by a second layer of SLG. This arrangement — two SLGs and the dielectric spacer — forms a SLG–insulator–SLG capacitor[70,72,77]. In both cases, silica cladding surrounds the capacitors. When a voltage is applied, carriers are driven into the Si waveguide core as well as into the SLG layer (Fig. 2a) or in the two SLGs (Fig. 2c), and the accumulation of carriers causes an $E_F$ shift[70–72,77]:

$$E_F(n_S) = sgn(n_S)\hbar v_F\sqrt{\pi|n_S|} \qquad (1)$$

where $v_F$ ~9.5 x $10^7$ cm $s^{-1}$ is the Fermi velocity and $n_S$ is the surface charge density. The voltage needed to accumulate $n_S$ is the sum of two contributions: the potential across the insulator, in both the Si–insulator–SLG and SLG–insulator SLG capacitors, and the quantum capacitance[78]:

$$|V - V_{DIRAC}| = \frac{qn_S}{C_{ox}} + K\frac{|E_F|}{q} = \frac{q}{C_{ox}}\frac{E_F^2}{\pi(\hbar v_F)^2} + K\frac{|E_F|}{q} \qquad (2)$$

where q is the electron charge, $C_{ox}$ is the oxide capacitance per unit area and $V_{DIRAC}$ is the voltage corresponding to the charge-neutral Dirac point, with K = 1 or 2 for single and double SLG, respectively.

The computed optical absorption in single or double SLG as a function of EF is shown in Fig. 2b,d. A commercially available mode solver is used to evaluate $n_{eff}$ and the optical absorption of the waveguide mode. SLG is modelled with an in- plane dielectric constant obtained from the optical conductivity[70], and the outof- plane dielectric constant is taken as equal to the graphite dielectric constant[79]. We use the closed formula for the complex optical conductivity[80]:

$$\sigma(\omega) = \frac{\sigma_0}{2}\left(\tanh\frac{\hbar\omega+2E_F}{4k_BT} + \tanh\frac{\hbar\omega-2E_F}{4k_BT}\right) - i\frac{\sigma_0}{2\pi}\ln\left[\frac{(\hbar\omega+2E_F)^2}{(\hbar\omega-2E_F)^2+(2k_BT)^2}\right] + i\frac{4\sigma_0}{\pi}\frac{E_F}{\hbar\omega+i\hbar/\tau} \qquad (3)$$

Here, $\sigma_0 = e^2/4\hbar$ is the SLG universal conductivity[81,82] and kBT is the thermal energy. T = 300 K is used in the simulations. Three values for τ are considered — 10 fs, 100 fs and 300 fs — for a free-space wavelength λ of 1,550 nm and $E_{ph}$ of 0.8 eV. When $|E_F| < E_{ph}/2$, photons may induce interband transitions[83]. This results in light absorption at rates as large as 0.1 and 0.2 dB μm$^{-1}$ for single and double SLG, respectively[71,72]. In this $E_F$ range, the absorption curves are almost independent of τ: interband transitions dominate over intraband ones[83]. When $|E_F| > E_{ph}/2$, interband transitions are forbidden as a result of Pauli blocking[83], and SLG would ideally be transparent. However, the smaller the τ, the larger the absorption, even when $|E_F| > E_{ph}/2$.

For Si-based waveguides composed of SLG (Fig. 2a), an $E_F$ shift is obtained by applying a voltage through the Si–insulator–SLG capacitor[75,77]. Carriers accumulate in the SLG and the underlying Si waveguide and, as a result of plasma dispersion[84] (Appendix 4), the carriers in the Si waveguide cause absorption. Thus, although carriers make SLG transparent for high $|E_F| > 0.5$ eV), they also make Si opaque. Instead, for small $n_s$ ($|E_F| < 0.4$ eV), the Si losses (green curve, Fig. 2b) are negligible, and absorption is mainly a consequence of interband transitions in SLG. When $|E_F| > 0.4$ eV, interband transitions in SLG are forbidden (Fig. 2). The black, blue and red curves, which represent the contributions to losses arising from SLG for different τ, have a similar behaviour in double SLG. Losses reach a minimum in the range $10^{-2}$–$10^{-3}$ dB μm$^{-1}$ for $|E_F| ≈ 0.5$–0.6 eV and then increase because of intraband transitions. However, losses in Si increase monotonically with $n_s$. The net result is a waveguide that is never as transparent. For Si-based waveguides covered with double SLG, losses decrease to minimum values of ~$10^{-2}$, $10^{-3}$ and 5 x $10^{-4}$ dB μm−1 for τ = 10 fs (black curve), 100 fs (blue curve) and 300 fs (red curve), when $|E_F| ≈ 0.5−0.6$ eV (Fig. 2d). If $n_s$ is increased further, so that $|E_F| > 0.6$ eV, losses increase again as a consequence of intraband transitions. Both single and double SLG electro-absorption modulators can be obtained by varying the absorption between $|E_F| < E_{ph}/2$ and $|E_F| > E_{ph}/2$ (refs[71,72]).

Equation 3 implies that a change in $E_F$ affects both the real and the imaginary part of the conductivity of SLG, more specifically the absorption, α, and refractive index, n. Therefore, SLG may be used to realize phase modulation[70,77]. In a MZI configuration, phase modulation allows for binary modulation formats, such as non-return-to-zero[85]. We demonstrated a SLG-based MZI phase modulator, transmitting over a standard fibre link[86]. This paves the way for a number of complex modulation formats for increased spectral efficiency in transmission systems, for example, phase-shift keying[87], differential phase-shift keying[88], quadrature phase-shift keying[89] and quadrature amplitude modulation[89]. All these are likely to increase the spectral efficiency of a communication link[90–92].

The computed Δneff as a function of bias in the Si–insulator–SLG and SLG–insulator–SLG capacitors are shown in Fig. 3a. We compute the voltage according to equation 2 by using a $SiO_2$ dielectric layer and compare the result with that of typical Si modulators[93–103] based on p–n junctions or capacitors in the same range of voltages. By biasing SLG to $|E_F| > E_{ph}/2$, where absorption is small, a neff modulation ~2 x $10^{-3}$ can be obtained[79]. At an operating wavelength of 1,550 nm, this accounts for a phase modulation ~8.1 rad per millimetre of propagation along the waveguide, over ten times larger than in typical Si phase modulation based on reverse- biased p–n junctions[59].

For comparison, we evaluate Δneff as a function of bias for a Si reverse- biased p–n junction modulator with p and n doping ~5 x $10^{17}$ cm$^{-3}$ (Fig. 3a). The modulation achieved by using two SLGs gives the largest Δneff, as well as the steepest Δneff. In the 2.5–4 V range, Δneff of double SLG as function of voltage is higher than that of other technologies, even though SLG is absorbing. Above $E_F$ ~0.45 eV (5 V), SLG becomes transparent, and Δneff has the steepest variation with applied voltage, and this allows for a short-phase modulation section.

In Fig. 3b, we compare the FOM of graphene modulators with those of p–n junction Si modulators, Si–insulator–Si modulators[104] (black curve) and InGaAsP membranes on Si (refs[105,106]). The points are experimental data taken from literature, whereas the lines are theoretical estimations for optimized modulator parameters. The Si–insulator–Si modulator comprises a vertical stack of a Si thin layer and a 5 nm-thick layer of $SiO_2$ followed by a layer of poly- Si. This Si–insulator–Si forms the waveguide core. Electrical contacts on the Si and poly-Si slabs form a capacitor across the insulating $SiO_2$ layer. By applying a voltage to the contacts, charge accumulates at the Si/$SiO_2$ and poly-Si/$SiO_2$ interfaces. Through plasma dispersion, these charges lead to phase modulation. Δneff is significant, because charges are accumulated in the middle of the waveguide, maximizing overlap of the optical mode.

Single or double SLG electro-absorption modulators can also be used as electro-refractive modulators, for example, in a MZI modulator with SLG phase shifters on its two arms[70,77]. The $FOM_{PM}$ for SLG- based electrorefractive modulators is about ten times higher than that for Si-based phase modulators when μ of SLG is high (τ > 100 fs), comparable to that of InGaAsP (refs[105,106]).

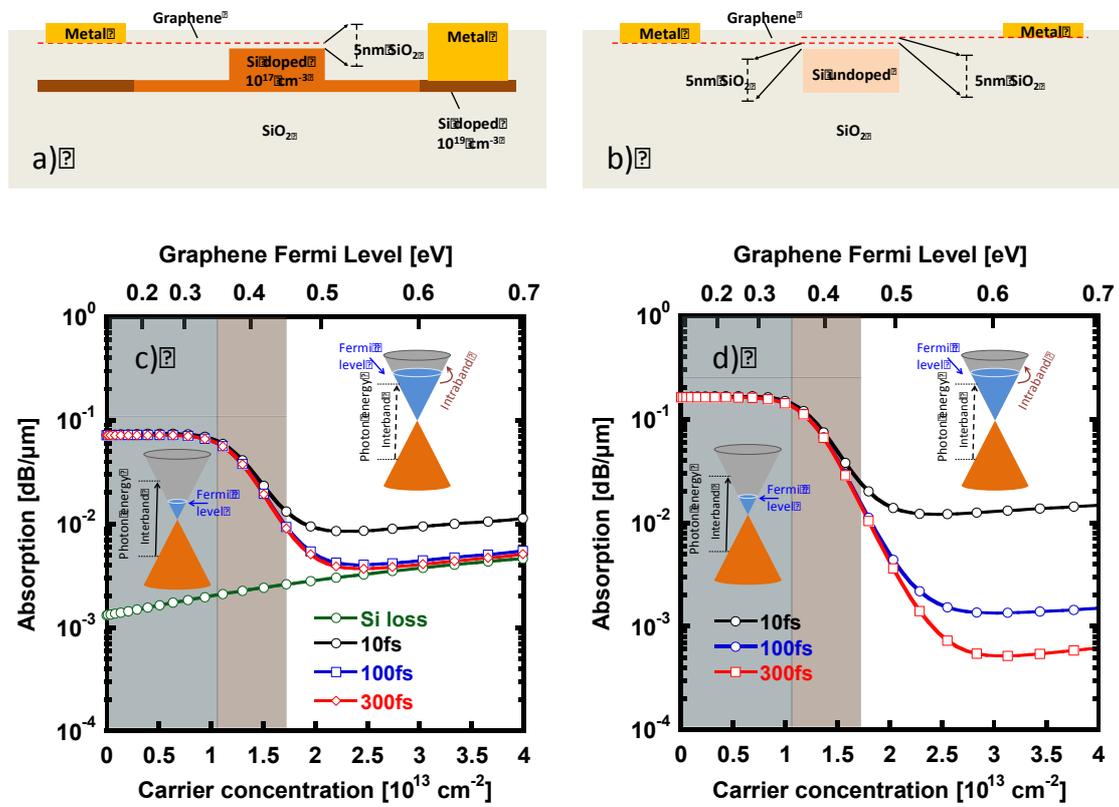

**Fig. 2 | Optical absorption profiles of SLG devices**. **a** | Schematic of a Si waveguide (with core dimensions of 480 nm (width) x 220 nm (height)) covered with one single- layer graphene (SLG). **b** | Calculated optical absorptions of single SLG devices (shown in panel a) with a scattering time, τ, of approximately 10 fs, 100 fs and 300 fs. The contribution to absorption from carriers in Si is shown in green. **c** | Schematic of a Si waveguide covered with two SLGs, namely, a double SLG device. **d** | Calculated optical absorptions of double SLG devices (shown in part b) with τ = 10 fs, 100 fs and 300 fs. In parts b and d, the blue zone denotes $E_F$ (Fermi energy) < $E_{ph}/2$ with $E_{ph}$ (energy of the propagating photons) ~0.8 eV for an operating wavelength of 1,550 nm. In this $E_F$ range, a photon impinging on SLG can be absorbed. When $E_F$ > 0.45 eV, absorption is suppressed as a result of Pauli blocking. In the pink zone, where 0.35 < $E_F$ < 0.45 eV, photons can be either absorbed or not, depending on $E_F$. Modulation of $E_F$ within the pink zone is used to obtain an electro-absorption modulator. In the white zone ($E_F$ > 0.45 eV), interband transitions are forbidden. However, intraband transitions can still cause variations in the dielectric constant of SLG. The white zone can be exploited to achieve phase modulation.

This higher $FOM_{PM}$ is a result of the combination of the large electro-refractive effect, $V_πL$, and low $α_{loss}$. A larger Δneff requires either a smaller voltage or a shorter L to obtain a π phase variation. For SLG modulators[70], $V_πL$ < 2.8 V mm (single SLG) and <1.6 V mm (double SLG) or <0.7 V mm for a double SLG embedded in the core of the waveguide[77]. In a typical Si p–n junction modulator, $V_πL$ ~20 V mm (ref.[101]), and in Si–insulator–Si capacitor modulators, $V_\pi L$ ~2 V mm (ref.[107]).

In p–n junction modulators or Si–insulator–Si modulators, $\Delta n_{eff}$ is due to the depletion or accumulation of charges (Appendix 4). However, charges absorb light[108], resulting in a large insertion loss. In the Si–insulator–Si modulator, poly- Si, which typically has a larger $\alpha_{loss}$ than crystalline Si, is part of the waveguide core. As a result, the losses in Si–insulator–Si[108] modulators and in depleted p–n junction modulators[109] are ~5 dB mm$^{-1}$ and 0.55 dB mm$^{-1}$, respectively. These limitations of Si–insulator–Si modulators — the large insertion loss and the presence of lossy poly-Si — are circumvented in SLG modulators. Therefore, in single SLG modulators, $\alpha_{loss}$ is comparable to depleted p–n junction modulators (Fig. 3b). In a phase shifter composed of SLG, a large $n_s$ ($E_F > 0.45$ eV at an operating wavelength of 1,550 nm) must accumulate in both SLG and Si to achieve carrier accumulation in the capacitor and, therefore, a considerable electro-refraction or electro-absorption effect. The doping of Si induces $\alpha_{loss}$ (see equation 12). Conversely, in the double SLG modulator, αloss can be small (Fig. 2d) because Si doping is not required (Fig. 2b). If the modulator is biased to an $E_F$ where interband transitions are inhibited, $\alpha_{loss}$ arises only from intraband transitions. For $\tau > 100$ fs, $\alpha_{loss}$ as low as 2 dB mm$^{-1}$ can be obtained[70]. In terms of the overall $FOM_{PM}$, double SLG modulators with $\tau > 100$ fs are at least 3–5 times better than InGaAsP/InP modulators[38,39], 5–10 times better than Si modulators and 10–20 times better than LiNbO$_3$ modulators[36,37].

For a double SLG device in the SLG transparency region, the theoretical estimate is $V_\pi L$ ~1.6 V mm for an undoped waveguide with double SLG[70], with an overall $FOM_{PM} < 2$ V dB for $\tau = 100$ fs and $FOM_{PM} < 1$ V dB for $\tau = 300$ fs (Fig. 3b). These theoretical values, if achieved, would result in devices that outperform Si photonic devices. However, hybrid technologies relying on III–V semiconductor membranes bonded on doped Si photonics waveguides can achieve performances similar to SLG photonics. For example, static phase shifters based on a 150 nm- thick InGaAsP membrane on 5 nm Al$_2$O$_3$ deposited on a Si waveguide showed $V_\pi L = 0.47$ V mm and an insertion loss of ~1.9 dB mm$^{-1}$ at an operating wavelength of 1,550 nm (ref.[106]). This modulation efficiency corresponds to $FOM_{PM} \approx 0.91$ V dB (ref.[106]). Similarly, an InGaAsP membrane bonded on 10 nm SiO$_2$ on an n-doped Si waveguide core has both high modulation efficiency and high speed[105]. With this configuration, a 3 dB modulation bandwidth of 2 GHz and a data rate of 32 Gb s$^{-1}$ with $V_\pi L = 0.9$ V mm were reported[105]. The InGaAsP–thin-oxide–Si stack exploits the capacitor concept first used in the Si–insulator–Si capacitor (SISCAP)[93], with the advantage of replacing the lossy poly-Si top layer with a highly efficient electro-refractive n- doped InGaAsP membrane[105,106]. The SLG–thin-oxide–SLG capacitor exploits the same effect, with good electro- refractive efficiency, but with the advantage that the capacitor is placed on a passive waveguide made of, for example, either Si, SiN or SiO$_2$ (refs[70,77]).

Plots of $FOM_{EA}$, defined as ER/IL, as a function of $n_S$ and $E_F$ for single and double SLG devices at different voltages are shown in Fig. 3c,d. The peak-to-peak voltage, $V_{PP}$, is the voltage needed to switch the modulator from a high to a low transmission level, or vice versa, and more specifically, to emit a '0' bit (smaller transmission) or a '1' bit (larger transmission). The resulting ER is the ratio between the high and low level transmitted power. $FOM_{EA}$ depends significantly on $\tau$: in a double SLG device for $V_{PP} = 2$ V, $FOM_{EA}$ ~5 for good-quality graphene ($\tau > $ ~100 fs), while $FOM_{EA}$ ~3 for poor-quality graphene ($\tau > $ ~10 fs).

To understand how these performances compare with the specifications of 100GE standards, we recall that the modulator for a 100GE CLR4 system is required to provide ER~4.5 dB, an IL~3 dB, therefore $FOM_{EA}$ ~1.5 (Appendix 2). SLG-based electro-absorption modulators can achieve these requirements in both single and double SLG configurations, even when $V_{PP} = 1$ V (Fig. 3c,d).

For comparison with SiGe electro- absorption modulators, let us consider a double SLG modulator with $V_{PP} = 2$ V and $FOM_{EA}$ ~5. This outperforms the Ge waveguide electro-absorption modulator in ref.[110] that operates at $V_{PP} = 2$ V, with ER~4.6 dB and IL~4.1 dB, corresponding to $FOM_{EA} = 1.12$. In another Ge- based electro- absorption modulator[111], $FOM_{EA}$ ~2–3.3 was reported for $V_{PP} = 4$ V and $\lambda = 1,610$–1,640 nm. Similar results were reported for GeSi electro-absorption modulators[112], revealing that double SLG electro-absorption modulators are more promising than GeSi modulators.

We note that SLG-based electro-absorption modulators operate across the $|E_F| = E_{ph}/2$ threshold, where interband transitions are excluded as a result of Pauli blocking[71,72]. In this condition, as a consequence of the simultaneous modulation of $\alpha$ and n, both amplitude modulation and phase modulation occur[113]. This results in an instantaneous variation in optical frequency, which is termed chirped modulation[85]. A positive chirp modulation is useful for transmission in anomalous dispersion fibre links to compensate for the highfrequency components of light, which travel faster than the lower-frequency ones[85]. The difference in velocity of the high-frequency and low-frequency components is caused by the negative chirp produced during propagation through the fibre link. This effect has been reported[113] to enable 10 Gb s$^{-1}$ transmission of light at $\lambda = 1,550$ nm along 100 km of standard single-mode fibre even with an input ER as low as 3.5 dB.

The capacitance of the modulator is the main parameter affecting the modulation speed of Si–insulator–SLG and SLG

insulator–SLG capacitors (Fig. 2a,c). Increasing the distance between the SLG layers improves the cut-off frequency of the capacitor, but increases the signal voltage and worsens both modulation efficiency and power consumption. These factors can be improved by embedding the SLG layers in the waveguide core[77], as well as by optimizing the length of the electrical leads (that is, the metal connections between contacts and SLG) and ensuring that the SLG length exceeds the width of the waveguide core.

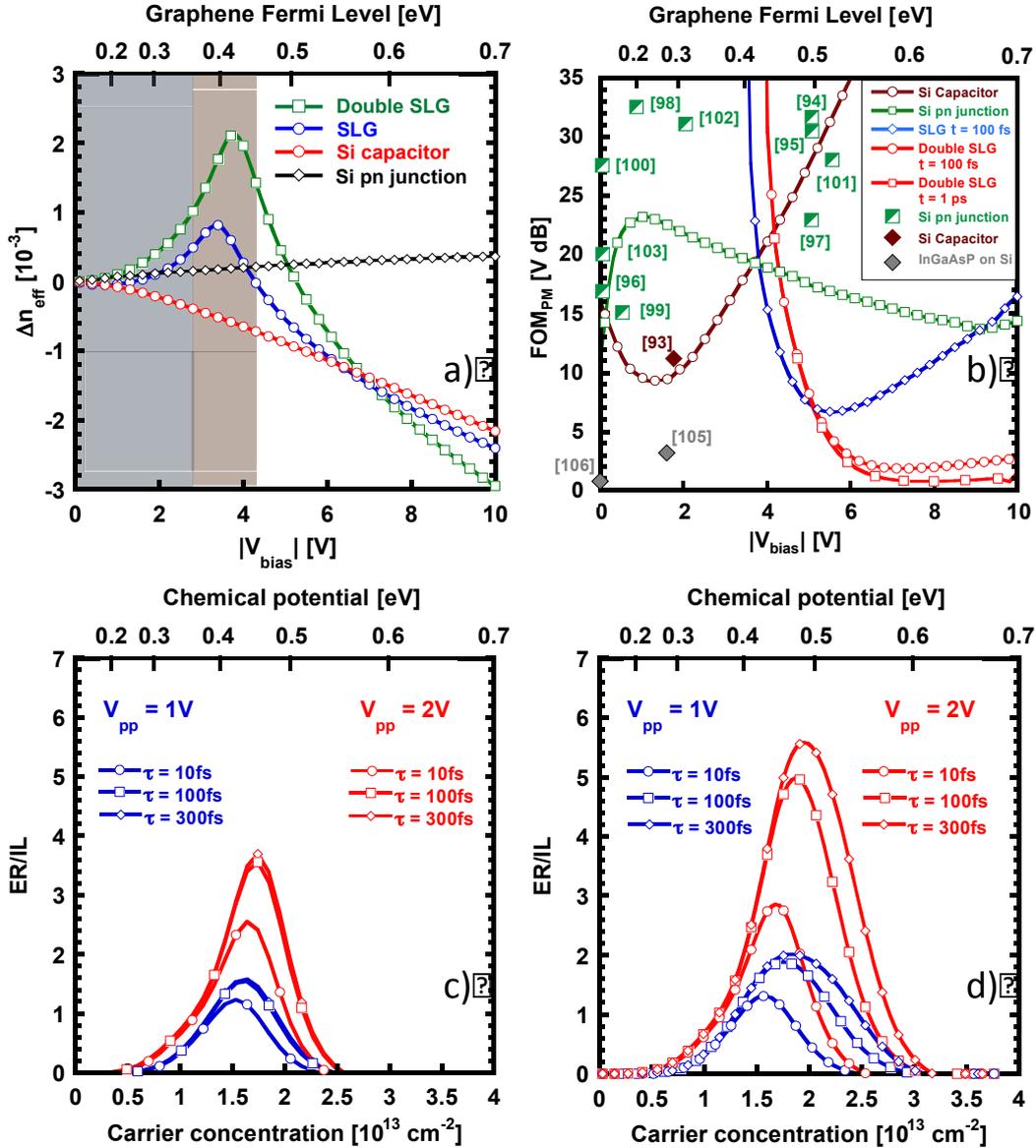

**Fig. 3 | FOM for electro-refractive modulators and electro-absorption modulators based on SLG. a** | Electro-refractive modulation in Si waveguides. The change in the effective index, Δneff, as a function of voltage for a mode guided by devices comprising either one SLG or two SLG covering the Si waveguides (green and red). Δneff as a function of voltage for a Si reverse- biased p–n junction modulator with a p and n doping of ~5 x $10^{17}$ cm$^{-3}$ (blue). Δneff as a function of voltage for a Si–insulator–Si capacitor modulator (black)[104]. The blue zone denotes $E_F < E_{ph}/2$, with $E_{ph}$~0.8 eV, for 1,550 nm. For these values of $E_F$, a photon impinging on SLG may be absorbed. The pink zone is the range $0.35 < E_F < 0.45$ eV, in which modulation can be used to achieve amplitude modulation. In the white zone ($E_F > 0.45$ eV), interband transitions are forbidden. However, intraband transitions cause a variation in the dielectric constant of SLG, achieving phase modulation. **b** | Figure of merit for a phase shifting functionality for a MZI electro-refractive modulator[86], $FOM_{PM}$, for one SLG on Si waveguides (green), two SLGs on Si waveguides (red), Si waveguide with a p–n junction (blue) and Si–insulator–Si capacitor waveguide (black) as a function of bias voltage, $V_{bias}$. The dots are experimental data from refs[93–103,105,106]. **c,d** | Figure of merit for electro-absorption modulation ($FOM_{EA}$) for increasing scattering time, τ, as a function of carrier concentration. Data for one SLG on a Si-on-insulator (SOI) waveguide are shown in panel c. Data for two SLGs on a SOI wire waveguide are shown in panel **d**. $V_{PP}$, peak-to-peak voltage. Panel b is adapted from ref.[86], Springer Nature Limited.

**Graphene integration on SiN**

The platform of Si photonics can be a base upon which processes or technologies are developed for integrated photonics — most importantly, transceivers. Modulation and detection in Si photonics are key elements of transceivers and require a layer of crystalline Si sandwiched between a buried $SiO_2$ insulator and a top cladding of $SiO_2$ (ref.[61]). This allows propagation of single optical modes in the waveguide (SOI wafer technology)[114]. The Si guiding layer can be p-doped or n-doped to enable modulation[61], and an epitaxial Ge layer can be grown on it to enable detection through carrier photogeneration[115]. A more convenient configuration is that of a pair of SLG layers, that is, a capacitor consisting of a SLG–insulator–SLG stack on a passive waveguide[72].

SLG–insulator–SLG capacitors have potential advantages over Si photonic modulators. First, the fabrication is not dependent on the waveguide material, electro-absorption modulation or electro-refractive modulation. Photogeneration is provided only by the SLG structures fabricated after the waveguide in postprocessing[116]. As a result, the optical guiding circuit does not require doping or Ge epitaxy for detection. Instead, it is fully passive, and only core material with no doping is required, which can be exploited to simplify the fabrication technology. Given this flexibility, SiN can be selected as the core material. SiN is amorphous, with a refractive index larger than that of silica ($n_{SiN}$ = 1.98 at 1,550 nm), and it is transparent in the visible and IR regions[117]. The large difference in the refractive index between the core (SiN) and the cladding ($SiO_2$) ensures mode confinement down to $\leq 1\mu m^2$ and therefore waveguide miniaturization[118]. To fabricate the waveguide, SiN is deposited on buried $SiO_2$ between SiN and Si. The SiN platform, compared with the Si photonics one, is low cost because it requires standard Si wafers rather than speciality wafers such as SOI. SiN single-mode waveguides are typically 1 μm wide at 1,550 nm (ref.[119]). This means that the lithography resolution is relaxed compared with Si single-mode waveguides, which are 0.5 μm wide[11]. The SiN core can be defined with either a low-resolution 400 nm node optical lithography stepper or with a mask aligner with a resolution <1 μm (ref.[120]). The cost of the low-resolution mask, compared with the high resolution one required for Si photonics, could be at least fivefold less[121]. The consequence of this reduction in cost is that the production volumes to amortize the investments are smaller, opening medium-volume (for example, telecom) and small-volume (for example, ultra-long-haul optical systems) markets. Thus, whereas manufacturing in a standard CMOS line leads to mass-market products, the graphene photonics approach permits the use of already amortized fabs and lowers the cost of fabrication, therefore opening medium-volume markets.

The subassembly, which integrates the SLG photonics chip, laser and fibre array, comprises the largest fraction of the total manufacturing cost[122]. The high cost is related to laser integration, fibre array coupling and pigtailing. In a SLG photonics circuit, by exploiting SiN as a passive waveguide platform, laser integration and fibre coupling are more fabrication tolerant than in Si photonics. Because the difference in the refractive index between core and cladding is smaller in SiN waveguides than in Si photonics ones, the numerical aperture of the SiN waveguide is closer to that of the laser or fibre to be matched, and the impact of the packaging is reduced. As a further example, a double SLG stack can be assembled on any other core material, such as silica[123] or other materials.

Si photonics transmitters are mainly based on MZI modulators[124]. Depending on the ratio between the bit time, $T_{BIT}$ (the time duration of a single bit), and the transit time of the optical wave through the modulator, $T_T$, modulators can be classified as lumped[70,77,93,94,105] or travelling wave[96–99,125]. $T_{BIT}$ is the reciprocal of the bit rate, defined as the number of bits per second. For example, in a transmission with a bit rate of 10 Gb s$^{-1}$, each bit has duration $T_{BIT}$ = 1/bit rate = 100 ps. For a lumped modulator, $T_T \leq T_{BIT}$, and for a travelling wave modulator, $T_T \geq T_{BIT}$ (ref.[126]). A MZI modulator is characterized by two electrical parameters: the total capacitance of the two optical phase shifters through which the voltage is applied and $V_{PP}$, which sets the extinction ratio of the optically encoded signal and, consequently, the required energy. The energy performance efficiency used in electronics, also called the power-delay product[127], here is called the energy cost per bit[128], corresponding to the energy associated to one bit, in pJ bit$^{-1}$ or mW GHz$^{-1}$. In a lumped configuration, an electronic circuit (driver) charges and discharges the total capacitance at $V_{PP}$ for each data transition[129]. The maximum power consumption of the two phase shifters is given by the energy consumption in a bit time[130]:

$$P_{out} = BR \frac{C_T V_{PP}^2}{2} \quad (4)$$

In a travelling wave configuration, the MZI modulator is driven by a terminated electrical transmission line placed on the MZI phase shifting arms[131]. The total capacitance is split into N capacitances along the line. The power required to generate an on–off modulation is[100,130]:

$$P_{out} = \frac{V_{PP}^2}{2Z_L} \quad (5)$$

where $Z_L$ is the characteristic impedance of the transmission line[100]. From equations 4 and 5, the energy cost (or energy per bit), expressed as $P_{out}$/bit rate, scales inversely with bit rate in the travelling wave configuration, whereas it is constant in the lumped configuration. The energy cost reduction for the travelling wave compared with lumped wave at the same $V_{PP}$ is $2C_TZ_L$ multiplied by the bit rate. For very large bit rates, for example, the 56 Gb s$^{-1}$ case discussed at the Optical Internetworking Forum OIF CEI-56G (Common Electrical Interface at 56 Gb s$^{-1}$)[132], this reduction can be relevant. If we consider the capacitance of a lumped Si modulator, $C_T \sim 1$ pF, and the load impedance of a travelling wave modulator[131], $Z_L = 50$ Ω at a bit rate of 56 Gb s$^{-1}$, then an approximately sixfold energy reduction is obtained. For this to be achieved using travelling wave operation, long-length (compared with radio frequency wavelength) modulators are required[109]. In this respect, SLG photonics outperform Si photonics. Phase modulation in Si photonics MZIs requires doping of the Si waveguides. Therefore, it is inherently lossy, with losses ~0.55 dB mm$^{-1}$ (ref.[109]) or 5 dB mm$^{-1}$ (ref.[107]). The losses in double SLG phase modulators in sections of MZI can be < 5 dB mm$^{-1}$ for $\tau > 100$ fs (ref.[70]) at $E_F$~0.5 eV (Fig. 2d).

This low loss in SLG can be exploited to increase the modulator length (Fig. 2b,d), allowing the travelling wave configuration. $V_\pi L$ for SLG is better than that for Si photonics MZIs. Thus, the signal voltage at a given device length is smaller, and this contributes to further reduce the energy cost per bit. A modulator for a 100 GBE interconnect with ER=8 dB (Appendix 2) can be achieved with $V_{PP}$~0.37$V_\pi$ (refs[124,133]). Assuming a double SLG lumped modulator, $V_\pi L = 1.2$ V mm (ref.[70]) and L = 400 μm (ref.[134]), $V_{PP}$ ~ 1.1 V. With $C_T$ ~1 pF (ref.[132]), we get 0.6 pJ per bit from equation 4, independent of bit rate and insertion loss. For comparison, a Si photonics modulator[132] has a consumption of 2–3 pJ per bit and IL~2.6 dB at 1,310 nm. If we consider a graphene-based travelling wave configuration with a device length of 1.2 mm and a bit rate of 56 Gb s$^{-1}$, $T_{BIT}$ ~$T_T$ = 17 ps and $V_{PP}$ ~0.35 V, the energy consumption is ~25 fJ per bit for $Z_L = 50$ Ω and IL~1 dB.

Double- SLG-based modulators have a better $FOM_{PM}$ (~1 V dB) than optimized Si photonics travelling wave MZI ($FOM_{PM}$ = 12 V dB)[125 as well as lumped Si photonics MZI modulators (3 V dB)[134]. The $FOM_{PM}$ of double SLG-based modulators is comparable to that of InGaAsP membranes on Si (ref.[105]). The performance of state-of-the-art Si photonics modulators and SLG-based modulators are compared in Table 2.

| Modulator Operating wavelength (T/E) | $V_{bias}$(V) /$V_{PP}$(V) | ER (dB) @Gb s$^{-1}$ | L (mm) | Total IL (dB) /PL(dB mm$^{-1}$) | $V_\pi L$ (V mm) | $FOM_{PM}$ (V dB) | Power (mW) | Energy (pJ bit$^{-1}$) @Gb s$^{-1}$ | Refs |
|---|---|---|---|---|---|---|---|---|---|
| Double SLG TWMZI: 1,550nm (T) | 6.5/0.3 | 3.0[a] | 1.3 | 1.56[b]/1.2 | 1.6 | 1.92 | 0.45 | 0.01@40 | 70 |
| Embedded (SiN) double SLG TWMZI: 1,550nm (T) | 5/- | - | 0.17 | 0.02[b]/0.1[c] | 0.7 | - | - | - | 77 |
| Double SLG lumped: 1,550nm (T) | 6.5/1 | 3[a] | 0.4 | 0.48[b]/1.2 | 1.6 | 1.92 | 12.8 | 0.32@40 | 70 |
| TWMZI: 1,550nm (E) | 0/0.36 | 3.5@40 | 2 | 12.5[d]/2.25 | 7.5 | 16.87 | 1.3 | 0.032 @40 | 96 |
| Lateral p-n junction: 1,540nm (E) | 5/6.5 | 6.5@40 | 3.5 | 3.85[d]/1.1 | 29 | 31.9 | - | - | 94 |
| PAM4 TWMZI lateral p-n junction: 1,300nm (E) | -0.5/2.16 | 6.0[e] | 2.8 | 5.0/1.02 | 14.7 | 15 | 135 | 4.8[e] | 99 |
| TWMZI lateral p-n junction: 1,300nm (E) | 0/1.5 | 3.4@40 | 3 | 5.5[d]/1.1 | 25 | 27.5 | 22.5 | 0.45@50 | 100 |
| TWMZI lateral p-n junction: 1,300nm (E) | 0/1.6 | 9.0@16 | 3 | 5.4/1.2 | 16.7 | 20 | 165 | 10.3@16 | 103 |
| TWMZI lateral p-n junction: 1,300nm (E) | -0.5/1.85 | 4.4@32 | 2.8 | 4.9/0.72 | 16.1 | 12 | 140 | 4.4@32 | 125 |
| TWMZI lateral p-n junction: 1,550nm (E) | -6/7 | 5.56@50 | 7.35 | 6.91[b]/1.04 | 26.7 | 27.8 | 245 | 4.9@56 | 101 |
| TWMZI lateral p-n | -5/6.5 | 7.5@40 | 0.75 | 1.2[d]/1.6 | 19.2 | 30.7 | 422.5 | 7@60 | 95 |
| junction: 1,550nm (E) | | | | | | | | | |
| TWMZI lateral p-n junction: 1,550nm (E) | -5/-3.5 | 7@56 | 3 | 9/1.3 | 18.5 | 24.05 | 57.5 | 1@56 | 97 |
| TWMZI lateral p-n junction: 1,550nm (E) | -1/1.6 | 3.1@40 | 3 | 6.2$^d$/1.2 | 27 | 32.4 | 25.6 | 0.64@56 | 98 |
| TWMZI lateral p-n junction: 1,550nm (T) | -1/4.2 | ->∞ | 0.9 | 4d/4.2 | 7.4 | 31 | - | - | 102 |
| SISCAP: 1,310nm (E) | 1.7/1 | 8.0@40 | 0.4 | 2.6$^b$/6.5 | 2 | 13 | 2-3$^f$ | 0.38 | 93 |
| III-V Si MOS capacitor: 1,550nm (E) | 0/-$^g$ | -$^g$ | 0.5 | 0.23$^b$/0.46 | 0.47 | 0.91 | - | - | 106 |
| III-V Si MOS capacitor: 1,550nm (E) | -1.5/3.5$^h$ | 3.1$^h$ | 0.25 | 1.0$^b$/2.6 | 1.2 | 3.12 | - | - | 105 |

**Table 2 | Performances of the main types of MZI** modulator. E, experimental; ER, extinction ratio; FOM$_{PM}$, figure of merit of the phase shifter section of the phase modulator; IL, insertion loss; L, device length; MOS, metal oxide semiconductor; MZI, Mach–Zehnder interferometer; PL, propagation loss; SISCAP, Si–insulator–Si capacitor; SLG, single-layer graphene; T, theoretical; TWMZI, travelling wave Mach–Zehnder interferometer modulator; V$_{bias}$, voltage bias; V$_{PP}$, peak-to-peak voltage; VπL, modulation efficiency. a) Achievable transmission rate depends on contact resistance of the SLG. b) Losses refer only to the phase shifting section of the device. c) Losses for E$_F$ ≈ 0.6 eV. d) Grating coupler losses not included. e) Four-level modulator at 28 Gigabaud per second, corresponding to 56 Gb s$^{-1}$. f) Consumption is for each Gb s$^{-1}$. Baud is the number of 'symbols' transmitted per second: in a binary system, bauds coincide with bits, while in a multi-level system using 2M different levels, each baud carries M bits[323]. g) Phase shifter characterized in static operating conditions. h_Cut-off frequency of 2.2 GHz in depletion mode.

**Graphene-based switching**

Ethernet-type interconnections in optical networks are based on IP packet switching to route data streams[135]. Packet switching can be performed in both data and telecom networks[135]. In telecom networks, this is achieved mainly in parts of the network where IP transfer is supported between access networks. The aggregated traffic of the different data streams is becoming sufficiently large (100 GBE and higher)[18] so that new functionalities can be performed by optoelectronic switches. At each switching node, opto-electrical and opto-electro-optical conversions are required to support electrical disaggregation, switching and re-aggregation, and aggregated packet routing[136,137]. In the overall aggregation and opto-electro-optical conversions, latency and energy consumption[138] are major issues. Si photonics technology can enable the switching of data streams, mitigating the bottleneck of latency and power consumption. Switching can be performed directly in optics, without resorting to opto-electro-optical conversion and without data packet handling[139]. This can be an advantage when large data streams are aggregated and switched. For example, power saving and latency reduction can be achieved when longer persistent data streams, like elephant flows, can occur in a datacom network[140]. An elephant flow is an aggregated data stream transmitted through a fixed path for a sufficiently long time (>10 s)[140]. In this case, the data stream can be routed along different optical paths (or different fibre links) with dedicated hardware (for example, an optical circuit switch)[141]. Other applications include periodic data centre backups to prevent and minimize data losses[142]; reconfigurable add–drop multiplexing of large data streams[143]; protection of a connection path through restoration and an alternative pre-provisioned path in both data and telecom networks[144]; implementation of network resource provisioning through software-defined networking[145], initially done for datacom, but now also suitable for telecom[146]; and reconfigurable fibre connections in a meshed photonic network, the so-called optical cross connection[144].

Another important field in telecommunications is wireless. Highly directive antennas will be used in 5G networks to steer the beam of a radio signal in the direction of several final users with high available throughput (up to 10 Gb s$^{-1}$ in the microwave or millimetre wave spectrum), also known as massive beam forming[147]. Beam-forming antennas may lead to new types of network that rely on more efficient resource allocation and optimized power consumption and require a more extended use of the frequency spectrum[148]. The interconnection of several radio base stations equipped with beam-forming antennas can be achieved in line with the 5G fronthaul and backhaul network evolution by using optical circuit switches[28]. As a consequence of the increased radio link bandwidth and the need for fast network responsiveness (<1 ms latency, depending on the service) with respect to previous generations, switching is set to become a pervasive optical function[25]. Using dedicated software that considers each single hardware network element as 'virtualized' (that is, representable as a controllable network entity characterized and/or summarized by some relevant tuneable parameters), optical switches will be able to handle aggregated 100 GBE and beyond data streams[19], being operated remotely on demand through all computational resources supplied by one or more interconnected data centres. This approach defines a smart photonic cloud network[149], capable of delivering an undetermined number of services directly to the network system and/or the final user. A high virtualization in optical switch operations can be performed by leveraging the potential of photonic integrated circuits, which can be efficiently monitored and controlled by software, further lowering the equipment cost per bit, the footprint and the power consumption[146].

Over the years, many approaches have been proposed for realizing optical switches[139,150], such as electro- optic (mainly in Ti:LiNbO$_3$)[151–155], acousto-optic[156–158], thermooptic[159–163], liquid crystals[164–167], microelectromechanical systems (MEMS)[168–173] and semiconductor optical amplifiers[174–179]. These devices require expensive equipment and could be replaced by Si photonics integrated circuits to comply with the requirements of miniaturization and cost-effectiveness[180], through mass manufacturability[181].

Si and SiN optical add–drop multiplexers (OADM)[182–184] have been realized to demonstrate switching[185–191]. OADMs are four-port filters in which a micro-ring resonator is placed between two waveguides and is coupled to the waveguides[183,184]. This component is designed for wavelength division multiplexed (WDM) optical networks, that is, systems that make use of multiple wavelengths (or colours) to transmit different data streams and increase the capacity of the system[192–194]. In the multiplexer, the waveguide that carries the incoming signal is usually referred to as the bus[183]. In a typical dense wavelength division multiplexed (DWDM) optical link[195], the bus carries 72 wavelengths between 1,520.25 nm and 1,577.03 nm (the so- called C- band)[196], spaced 100 GHz apart. The second waveguide of the multiplexer is referred to as the drop[183]. When the incoming light has a wavelength that coincides with the resonance of the micro-ring resonator, light is transferred from the bus to the drop. This operation mode is called DROP[184]. When the incoming light is out of resonance, it proceeds unswitched, and the operation mode is called THROUGH[184].

DROP–THROUGH operation is determined by coherent interference inside the ring[183]. The light circulating in the micro-ring resonator interferes at each round trip with the incoming light from the THROUGH channel. At each round trip, a fraction of the field is extracted from the ring in the DROP channel, while the remaining part continues in the micro- ring resonator[183]. This operation implies a latency that, as in any resonant circuit, grows with the quality factor of the resonance. However, the quality factor cannot be too high or the spectrum of the signal travelling inside the ring becomes heavily distorted. Off- resonance, the periodical interference cancels out, and the field from the incoming bus waveguide continues in the THROUGH channel with no DROP. Switching configuration (that is, the possibility of enabling /disabling the DROP of each single wavelength in a DWDM link) is increasingly important in optical networks[193]. In a micro-ring resonator, this corresponds to tuning and/or detuning the resonances with respect to the wavelength of the signal that has to be dropped. For a switch to drop a single wavelength, the micro- ring resonator free spectral range (FSR) needs to be greater than the WDM spectrum to avoid spurious drops[197]. Detuning the resonance of the micro-ring resonator can be achieved by metal heaters[198–202] in the interlayer dielectric (waveguide cladding oxide) on top of the waveguide. Electrical energy is supplied to the heaters, which diffuse heat to the waveguide[202]. The thermo-optic effect[203,204] results in an increase in the index of refraction of Si with temperature[205], leading to the detuning of the micro- ring resonator (that is, a shift in resonant frequency)[202].

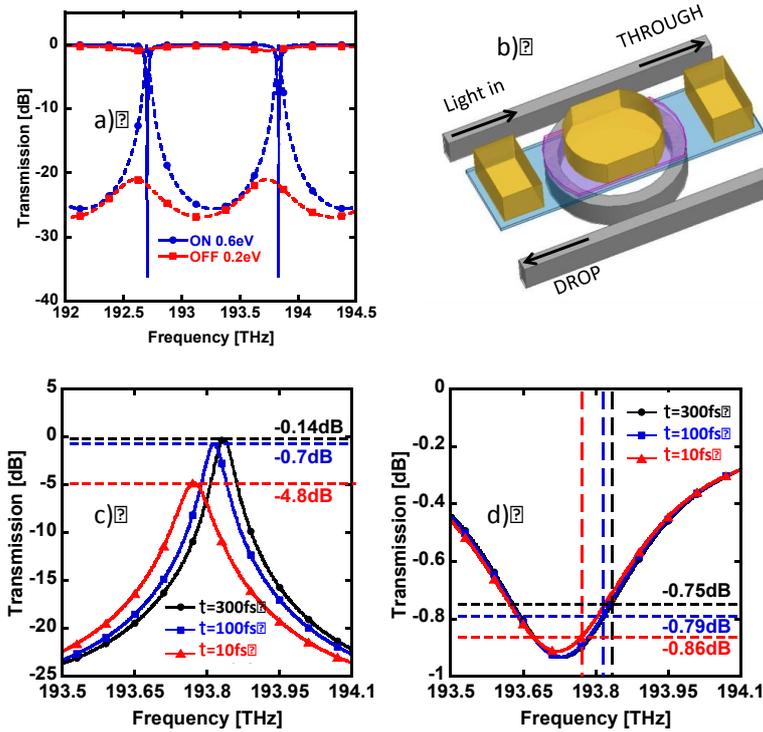

**Fig. 4 | transmission of a reconfigurable optical ADD–DROP multiplexer and effect of τ on its performance. a |** Transmission spectra as a function of frequency for a Si micro-ring resonator in the ON and OFF states (free spectral range (FSR) = 1.2 THz, bandwidth = 20 GHz and τ= 300 fs). In the ON state (minimum micro- ring resonator absorption), light circulates inside the ring and, at resonance, a coherent superposition of waves accumulates at the DROP port. In the OFF state (maximum micro-ring resonator absorption), light cannot travel through the ring, and it is directed to the THROUGH port. **b |** Schematic diagram (not to scale) of the proposed switch, realized by coupling two Si or SiN waveguides (grey) to a micro-ring. On top of the micro-ring, two layers of single-layer graphene (SLG) (blue and pink) are used to modulate losses in the micro-ring. **c |** Transmission spectra of DROP in the ON state as a function of frequency and τ. **d |** Transmission spectra of THROUGH in the OFF state as a function of frequency and τ. The vertical dashed lines represent the position of the resonances in the ON state. The horizontal dashed lines are added to facilitate reading peak values of DROP in the ON state (part c) and THROUGH in the OFF state (part d) at the same wavelengths of the peaks of DROP in the ON state.

SLG may represent a breakthrough in switching thanks to a new approach that exploits high-μ SLG electro-absorption rather than tuning. By placing two SLGs on the micro-ring resonator waveguide and by changing the voltage applied to those SLGs, losses in the micro-ring resonator can be varied from very large (~1,000 dB cm$^{-1}$) to small (<10 dB cm$^{-1}$)[79]. When SLG losses are large, the field circulating in the microring resonator is entirely absorbed in a single round trip, the interference with the incoming signal is suppressed, and DROP is disabled. When the loss is negligible, light can resonate and DROP is enabled. The main difference between this SLG- based switch and the Si photonics counterpart is that enabling and/or disabling of the DROP is obtained by suppressing, rather than detuning the micro-ring resonator resonances[206]. As a consequence, SLG-based switching has a major advantage over its Si photonic counterpart. In Si photonics, DROP disabling is obtained by placing the resonance between two adjacent channels in the ITU-T (Telecommunication Standardization Sector of the International Telecommunications Union) DWDM grid. In this way, the crosstalk between channels, defined as the ratio between the power of spurious and main signals, becomes critical, and the system tolerance becomes tight (that is, all the wavelengths in the transmission system should be locked with high accuracy to the ITU- T frequency grid)[207–212]. Another advantage of a SLG switch is the power consumption. The usual scheme based on thermal tuning of the Si micro- ring resonator leads to a continuous power consumption ~0.11 nm mW$^{-1}$ (ref.[213]). Assuming a micro-ring resonator resonance trimming due to fabrication errors ~1 nm, this corresponds to ~9 mW for each ring[213]. In the SLG switch, the variation in losses for resonance suppression is obtained by capacitive charging[71,72]. Thus, there is no power consumption in static operation, which is the normal state for a switch. Energy is consumed only during capacitance charging[206]. A further advantage is that the functionality that depends on SLG is part of post-processing and is therefore independent of the waveguide platform.

The performance of a SLG-based switch is shown in Fig. 4a. This device, which is composed of the waveguides depicted in Fig. 2c, is schematically represented in Fig. 4b. The transmission spectra in the C-band for a four-port micro-ring resonator (radius of 10 μm) with a FSR of 1.2 THz and a bandwidth of 20 GHz (at 1 dB) are shown (Fig. 4a). The device has an ADD DROP filter suitable for WDM switching[79]. The isolation from adjacent channels is achieved with τ ~300 fs (Fig. 4c). The performance in terms of suppression of resonance and, hence, DROP disabling, is determined by the maximum absorption achievable in a round trip along the micro-ring resonator. As a result, the double SLG configuration is preferable to the single SLG configuration[71,72]. DROP requires the coherent interference of signals travelling within the micro-ring resonator[183]. This is the ON state, and the losses in a single round trip must be as small as possible. SLG needs to be transparent for DROP to be performed. Similar to the phase modulation case, this transparency is related to τ. DROP spectra in the ON state can give a quantitative estimate of the influence of τ on the performance of the switch (Fig. 4d). For τ ~10 fs, the loss of SLG is large (>$10^{-2}$ dB μm$^{-1}$) at any value of EF (Fig. 3d). In a 10 μm ring, this accounts for a loss ~0.6 dB per round trip. This limits the coherent superposition of signal replicas inside the micro-ring resonator to ~20 waves. The result is a ~4.8 dB loss in the DROP channel (Fig. 4c). A larger τ gives rise to a smaller $α_{loss}$ (Fig. 2d) ~1 x $10^{-2}$ for 100 fs and ~5 x $10^{-3}$ dB μm$^{-1}$ for 300 fs. Therefore, a coherent superposition of a greater number of replicas inside the micro-ring resonator (~200 waves and ~1,000 waves for ~100 fs and 300 fs, respectively) gives a much smaller loss in the DROP channel ~0.7 dB and 0.15 dB for ~100 fs and 300 fs, respectively. For the THROUGH channel, an insertion < 1 dB is obtained when operating the ring in the OFF state, when setting $|E_F| < E_{ph}/2$, with $E_{ph}$ ~ 0.8 eV in the C-band (Fig. 4d). Losses as large as 0.1 dB μm$^{-1}$ may arise because of allowed interband transitions (Fig. 2d) with an almost negligible dependence on τ. Incoming light from the bus is partially coupled to the ring, the light is rapidly absorbed in the micro-ring resonator at a rate ~6 dB per round trip, and no coherent interference is formed. As a result, the micro-ring resonator DROP channel is disabled, and the light travels to the THROUGH port.

**Graphene photodetectors**

Graphene photodetectors have been extensively reviewed[49,51,53] and integration of graphene-based detector arrays with CMOS electronics was demonstrated[214]. Here, we focus on waveguide-integrated and speedoptimized photodetectors relevant to Si photonics. $R_{ph}$ should be comparable to that of Ge devices in Si photonics, that is, ~0.85–1.15 A W$^{-1}$ at an operating wavelength of 1,550 nm (ref.[41]).

The key principle of photodetectors is the conversion of absorbed photons into an electrical signal. Several detection mechanisms have been identified for graphene photodetectors, including photovoltaic[215,216], photo-thermoelectric[216,217], bolometric[218], photogating[219] and plasma-wave-assisted[220]. Each of these may become dominant in different photodetector configurations, such as SLG p–n, metal–SLG and single-double–SLG junctions[216,221]. SLG has unique properties and advantages for photodetectors[49,53]. For example, SLG has a zero-gap band, which implies frequency-independent absorption[69], allowing for light detection from the UV to far-IR region with a single material[222] and in a single technological step. Another advantage is the potential for less than picosecond photovoltage generation[221] and fast photo-switching rates, thus far up to ~270 GHz (ref.[56]). SLG also has high internal quantum efficiency with a ratio of electrons produced per photons absorbed >80%[42].

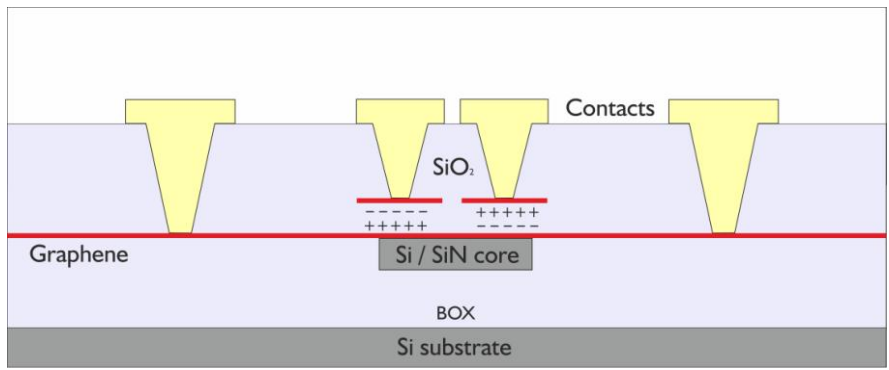

**Fig. 5 | Double-gated thermoelectric photodetector.** The lower SLG provides thermoelectric generation and voltage detection. The upper SLGs gate the lower SLG to a working point in which the Seebeck coefficient is maximized. The two independent SLG gates induce a carrier gradient in the middle gap. Light travelling in the waveguide is absorbed by the lower SLG. The upper layers of SLG are sufficiently distant from the propagating optical mode to avoid absorption. Photogeneration in the lower SLG maximizes the photovoltage. BOX, buried oxide.

In SLG, the photo-thermoelectric effect can be optimized for both high detection speed (25–100 GHz) and efficiency[223], potentially above 100% as a result of hot carrier multiplication[224]. The photo-thermoelectric effect involves the following steps: first, electron–hole pairs are generated by the absorption of photons, and second, ultrafast carrier scattering generates hot electrons and holes on a <50 fs timescale[224,225]. These electrons and holes generate a local photovoltage via the Seebeck effect[226], which drives a current from source to drain. When the light is switched off, electrons cool back to equilibrium in 2−4 ps (refs.[225,227,228]).

This cooling time yields the intrinsic limit of the photo-switching rate. The photo-thermoelectric effect can be highly efficient, as a large fraction of the photon energy is captured as electron heat owing to the ultrafast carrier scattering and weak coupling to phonons[222,225]. This results in a voltage source, because the Seebeck effect is an electromotive force that generates a voltage rather than a current[226], which can be used in transceivers needing a voltage to drive the receiver electronics. In contrast to a photovoltaic detector, in which the photogenerated current is typically amplified and converted into a voltage by a transimpedance amplifier[228], a photo-thermoelectric-based graphene photodetector generates voltage directly[225], and the transimpedance amplifier can be replaced by a simpler voltage amplifier. This property has advantages in terms of cost reduction and power consumption. In addition, direct voltage detection may overcome issues with the dark current of photovoltaic schemes with bias.

The photo-thermoelectric effect has been exploited extensively for graphene-based photodetectors[55,57,229–234]. On-chip integrated photodetectors with Si photonics have been reported[232–236], typically based on metal–SLG–metal structures evanescently coupled to Si waveguides. In these photodetectors, the guided mode enables longer interaction between SLG and the optical waveguide compared with free-space illumination. This longer interaction raises the optical absorption above 2.3% and, by increasing the interaction length, up to almost 100% of the light is absorbed and can contribute to photovoltage. Because of the evanescent coupling, the typical length needed to achieve nearly complete absorption in metal–SLG–metal photodetectors is ~40–100 μm. A speed-optimized graphene photodetector with a rate of ~50 Gb s$^{-1}$ was reported[237]. The device consisted of a chemical vapour deposition (CVD)-grown SLG on a Si waveguide operating at 1,550 nm. The evanescent field of the mode propagating in the Si waveguide overlaps with a p–n junction as a consequence of the $E_F$ shift at the metal interface of one of the contacts. One limitation is the contact metal in the evanescent field of the optical mode, which leads to a reduction in $R_{ph}$. The small size gave a capacitance of ~20 fF and a resistance of ~185 Ω, with a large bandwidth[237].

The highest $R_{ph}$ photodetectors to date are ~40 μm in length and integrated on a 520 nm-wide Si waveguide[42]. The drain and source were placed asymmetrically with respect to the waveguide core to create a p–n junction near the optical mode and obtain a net voltage drop. A gate was also used to maximize the Seebeck coefficient of SLG. SLG was encapsulated in hexagonal boron nitride (hBN) to improve μ to ~40,000 cm$^2$ (V$^{-1}$ s$^{-1}$) and the Seebeck coefficient, resulting in a larger $R_{ph}$~0.36 A W$^{-1}$. The side contacts provided a resistance as low as 77 Ω. The 3 dB bandwidth response was 40 GHz, very close to the target value for an optical communication receiver[238].

These results are promising for optical communication links, but there is a drawback: upon application of a bias, a continuous current flows. However, if two SLGs are arranged as shown in Fig. 5, the lower SLG provides thermoelectric generation and voltage detection, while the upper SLG acts as a split-gate, tuning the lower SLG to the optimized Seebeck coefficient. Calculations based on realistic parameters, for example, a Seebeck coefficient ~0.2 mV K$^{-1}$ (feasible for high-quality SLG with μ > 10,000 cm$^2$ V$^{-1}$ s$^{-1}$) with very low ($E_F \leq 40$ meV) residual charge density and a cooling thermal conductance ~70 nW K$^{-1}$ m$^2$ (ref.[226]) predict $R_{ph}$ > 0.8 A W$^{-1}$ or, in case of voltage detection, that is, the measurement of the electrical output in volts per optical input (in Watts), $R_{ph}$ > 100 V W$^{-1}$.

Another important performance metric of photodetectors is the normalized photo- to dark- current ratio, NPDR = $R_{ph}/I_{dark}$. The higher the NPDR, the larger the photodetector noise rejection and the ability to perform when interference (noise) is present. To achieve higher NPDR, $I_{dark}$ must be reduced, and $R_{ph}$ must be maximized. A promising route to increase $R_{ph}$ while minimizing $I_{dark}$ is to create a Schottky junction with rectifying characteristics (that is, a diode) at the SLG–Si interface[239]. By operating a Schottky diode in reverse bias (photoconductive mode), $I_{dark}$ is suppressed compared with $R_{ph}$, and the entire Schottky contact area contributes to photodetection. A compact (5 μm in length), waveguide-integrated, plasmonic-enhanced metal–SLG–Si Schottky photodetector was reported to have $R_{ph}$ ~0.25 mA W$^{-1}$ at 1.55 μm (ref.[239]). When the same detector is reverse biased with 1 V, $R_{ph}$ becomes ~85 mA W$^{-1}$, and Idark becomes ~20 nA (ref.[239]). This detector configuration shows a one order of magnitude increase in $R_{ph}$ over that of the standard metal–Si configuration without SLG. By taking advantage of the Schottky diode operation in the reverse bias, $R_{ph}$ could be further increased[239] up to ~0.37 A W$^{-1}$ at 3 V, comparable to that of state-of-the-art SiGe devices[41].

**Wafer-scale integration**

Most devices reported to date are on the laboratory scale and have contacts fabricated using metal lift-off[240]. This is not suitable for the very large-scale integration required for modern chip manufacturing, because liftoff has limitations, such as redeposition of metal, formation of ears at the metal edges and partial retention of the metal[240,241].

A standardized SLG–CMOS-compatible contacting scheme is yet to be developed. Studies reporting full integration of the wafer and SLG[242–245] are limited to examples in which the SLG is integrated at the last level of integration[242] or combined with metal contacts through lift-off[243–245]. This limitation hinders the adoption of SLG technology by the semiconductor industry. A SLG–CMOS- compatible integration module consisting of a sequence of processing steps in conventional CMOS tools, which guarantees compatibility with the reliability standards of the semiconductor industry, is needed to persuade industry to adopt SLG as a viable and reliable alternative to conventional materials.

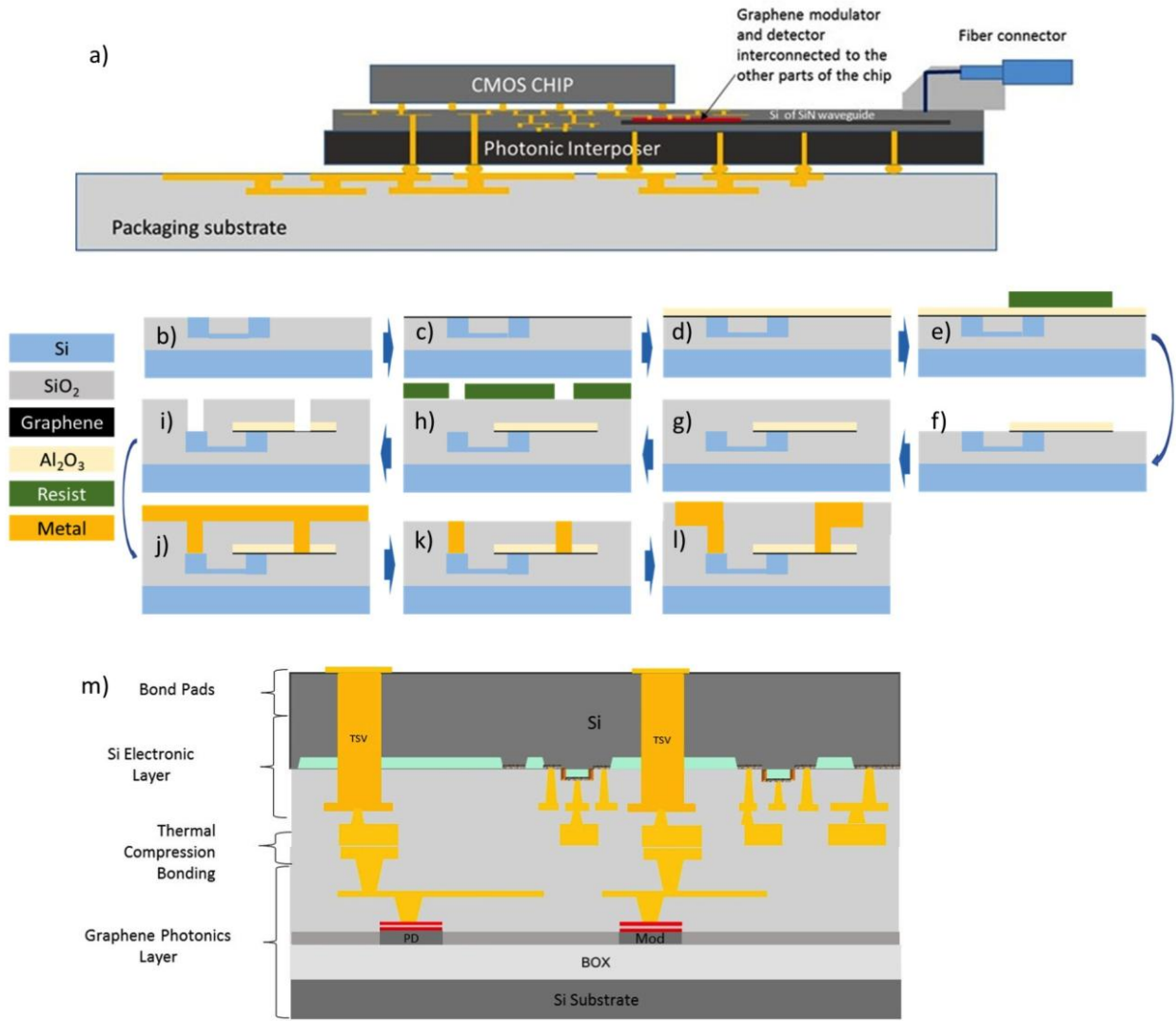

**Fig. 6 | Process flow of a SLG photonics integrated device. a** | Schematic of a single-layer graphene (SLG)-based transceiver integrated on a Si photonic interposer. The Si interposer was introduced by Xilinx in the 2.5D integration[11] to connect multiple chips sideby-side and provide high- bandwidth connections between dies, thus redistributing the die map to the packaging. The devices are interconnected with other components through a Cu interconnect back-end-of- line[241]. The interposer is connected to the packaging substrate using through- silicon vias (TSVs)[322]. **b** | Fabrication steps for a planar optical waveguide. **c** | Integration of a complementary metal oxide semiconductor (CMOS) circuit on a SLG photonics circuit. TSVs are thermally bonded at the interface between the SLG photonic layer and the electronic layer. The transmission from the driver circuit and the connection to the detection circuit are provided by the low power consumption, short interconnections with TSVs. Mod, modulator; PD, photodetector.

To allow the integration of SLG-based devices heterogeneously packaged with Si technology (Fig. 6a), SLG should be integrated through modules similar to those used to integrate semiconductor devices[246].

At present, the connections between devices in semiconductor applications are typically achieved by Cu damascene modules[247]. In this process, developed by IBM[248] and Motorola[249], the dielectric is patterned by dry etching with trenches or vias in which the conductor metal is later deposited (Fig. 6b). On the dielectric, a metal that overfills the trenches is deposited. Then, chemical mechanical polishing is used to remove the overburden metal on top of the dielectric. This process is known as 'damascene', by analogy to the art of incrusting wires of Au (and sometimes Ag or Cu) on the surface of Fe, steel or bronze[247]. The integration of a SLG device on a waveguide contacted through a damascene module is described in the sequence shown in Fig. 6b. The initial six steps describe SLG integration on a waveguide, and the remaining steps are related to the damascene contact module[248,249].

The fabrication does not require complex implantation sequences to activate the semiconductor locally. In addition,

there is no need for seeding layers to epitaxially grow crystalline materials on the waveguides for light detection or modulation, in contrast to the fabrication of Ge photodetectors[41,250] and SiGe or Ge electro-absorption modulators[60,251]. SLG integration reduces the complexity and the number of steps compared with 3D device integration[60,251], as well as the total temperature budget to integrate the active detector or modulator on the Si waveguides. A reduced temperature budget decreases the threat to the integrity of the devices processed earlier on the exposed wafer[252–254] and allows for flexible hetero-integration. The ease of integration of SLG devices with Si, as well as the reduced footprint of the SLG devices, should afford photonics technology increased cost-effectiveness compared with the competing semiconductor solutions.

To assess the costs of SLG–CMOS integration, progress in the integration maturity is needed. The fabrication steps in Fig. 6b require engineering to control performance and reproducibility and to achieve compatibility with the standards of the semiconductor industry. SLG can now be grown with similar quality to exfoliated SLG on metal or metal- coated substrates[255]. Growth of 5 cm x 50 cm SLG with >99% oriented grains is possible[256] for 300 mm wafers. Direct growth of SLG on Si wafers (hence, avoiding a transfer step) has been realized[257]. However, this method will not be adopted by the semiconductor industry because it requires wafers with processed devices. The highly diffusive catalytic metal (in this case, Cu)[255] is brought into contact with the dielectric of the target wafer and will diffuse into it when exposed at the temperature at which SLG growth occurs (at least 300 C)[258]. This temperature will destroy the integrity and performance of the dielectric and/or lifetime of the devices[259,260]. The present understanding is that the growth of high-quality SLG requires a metal catalyst and high temperature, as well as an efficient separation of SLG from the metal catalyst after growth[255] followed by transfer to the target wafer[261,262] (Fig. 6b). The manipulation of SLG is one of the most critical steps for SLG integration on Si and should be done in a controlled environment during transfer. SLG is an impermeable layer for molecules[263]. An uncontrolled environment results in contamination at the transfer interface, leading to uncontrolled traps and random strain fluctuations, one of the dominant sources of disorder in SLG devices[264]. Transfer has the advantage of allowing the interface between SLG and the target wafer to be engineered. Charge traps at the interface between the oxide and SLG result in a distribution of positive and negative doping puddles[265], which affect the local $E_F$, resulting in non-uniform electro-absorption[75] and μ reduction[266], with decreased device performance. The transfer of SLG to the target surface should be engineered to secure reliable adhesion. SLG has no dangling bonds on the surface to chemically interact with the surrounding dielectric, and the adhesion to the dielectric substrate is secured through van der Waals forces[267].

SLG encapsulated between hBN flakes enables a high μ at room temperature[251,268,269]. hBN has an atomically flat surface that significantly reduces electron–hole puddles compared with $SiO_2$ (refs.[270,271]). This suggests that a fundamental step towards control of the performance of large-area CVD-grown SLG can be achieved through the integration of SLG sandwiched between hBN. A single layer of wrinkle-free hBN can be grown on sapphire[272], paving the way for the integration of engineered heterostacks on the wafer scale[273]. The growth of a dielectric on SLG (Fig. 6b) without affecting the SLG opto-electric performance and avoiding defect formation and/or chemical interactions is another challenge. Plasma assisted dielectric deposition technologies tend to induce defects in SLG[274,275]. The atomic layer deposition (ALD) of high-k dielectrics has been studied[276]. The role of the nucleation density in order to achieve rapid layer closure has been extensively investigated277,278. It is imperative for the starting surface to provide enough reactive sites for reactions with the ALD precursors277. On selfpassivated materials, the nucleation of the dielectric typically occurs at the reactive defect sites. Therefore, it is possible to correlate the efficiency of the closure of the ALD layer to the quality of the SLG, where a less effective closure is observed for high-quality material, as the nucleation density is linked to the number of defects[279,280]. The state of the art is non-robust for future material improvement. Therefore, alternative seeding approaches need to be developed to realize the growth of the dielectric independent of SLG quality.

A further required development relates to the metals incorporated as SLG contacts (Fig. 6b). Some metals are not compatible with the CMOS production environment (as detailed in the International Technology Roadmap for Semiconductors) because their use affects device reliability and yield. Typically, only Al, W, Cu, Ni, Co, Mo, Ti and Ta are compatible with CMOS fabrication[281]. The contact architecture to SLG also has an impact on the complexity of the integration scheme, and edge contacts have proved to be the optimal architecture[282] and the easiest to achieve in a damascene module.

The performance gains and cost savings should push the industry to invest in the integration of graphene and related materials in the Si production line. Once answers are developed for these challenges, integration will comply with the standards of the semiconductor industry and will pave the way for the adoption of technologies based on graphene and related materials as a standard in the portfolio of the semiconductor industry. An alternative approach is based on single- crystal transfer in a predetermined position[283]. This involves growing single-crystal SLG at nucleation points predefined on a Cu support. These single crystals are set to overlap with the device on the destination wafer after transfer. The advantages are as follows: first, the growth of multiple individual single crystals is less challenging than growing a single wafer-size crystal; second, the predetermined position ensures the transfer of SLG crystals only where needed by design (for example, onto the waveguide modulators and detectors); third, transfer printing can be used to populate an entire wafer[284].

Another consideration is the integration of the electronics (electrical driver and transimpedance amplifier, TIA) and photonics on the SLG-based photonic circuit. The electronic integrated circuit (EIC) wafer and the optical integrated circuit (OIC) wafer (Fig. 6c) can be integrated by thermally bonding contacts or by using Cu pillars. This is important because the SLG photonic layer (OIC) consists of a post- processed SLG stack on passive guiding structures. SLG post-processing is therefore the final stage in the fabrication of the photonic layer. Contacting and metallization are achieved in the back-end-of- line, with no perturbation in the SLG quality or optical or electrical characteristics. This ensures full compatibility of SLG photonics with the electronic circuitry in the integration process.

**Conclusions and outlook**

The telecom and datacom industries are driven by the continuous increase in requirements for the bandwidth of communications. The advent of the 5G communication era will boost the bandwidth requirements as a result of the introduction of communications with the world of high- definition virtual reality, augmented reality, gaming, as well as connected objects, more specifically, the IoT. Managing the increased bandwidth demand requires significant advances in photonics hardware, well beyond incremental improvements.

We consider graphene-integrated photonics for telecom and datacom systems as an evolutionary step in integrated photonics and, in particular, Si photonics. The main advance enabled by the adoption of graphene is post-processing on a passive waveguiding structure, dedicated only to passive optical circuitry. All active functionalities, such as modulation, detection and switching, are graphene-based and applied on the passive underlying optical circuit. The separation of the guiding circuit from active functionalities leads to a technology that does not require a full integration approach, unlike Si photonics, which is strongly tied to CMOS processing.

We analysed the main functions (modulators, detectors and switches) of graphene photonics and compared them with established technologies. The graphene photonics modulator exploits carrier accumulation in a SLG–insulator–SLG capacitor and can provide the necessary bandwidth to match the telecom roadmap evolution, combined with low energy consumption (0.1 pJ bit$^{-1}$), size miniaturization (phase shifter length <0.5 mm or electro-absorption length <0.1 mm) and, most importantly, a $FOM_{PM}$ ~0.1 V dB for the phase shifter, which represents a significant improvement with respect to existing technologies, combined with a $FOM_{EA}$ of ~3 dB for the electro-absorption case. Graphene phase shifters can be driven with a voltage ≤1 V. This is relevant to the cost of the transmitter and its power budget, as the signal to the modulator is usually supplied by an electronic driver. If the voltage is ≤1 V, it is possible to design simplified electronics with no need for a specific electronic driver.

At present, graphene bolometric detectors provide responsivity of at least ~0.5 A W$^{-1}$ (ref.[285]), while those based on the photothermal effect have at least ~0.4 A W$^{-1}$ or 10 V W$^{-1}$ responsivity, which is bound to increase with mobility optimization, combined with negligible dark current. Such detectors can be used in either current or voltage mode. The conventional configuration is based on current detection and requires electronics, such as a transimpedance amplifier, for current- to-voltage conversion and amplification. However, the photothermal effect in graphene generates a voltage; therefore, the voltage configuration is more natural for graphene. The receiver design is thus simplified and is less costly to produce.

Graphene-based modulation and detection are key elements in a point-to-point transmission system. A telecom or datacom system also requires optical switching to route the signals through the communication network. The graphene-based switch is a building block that, analogous to a modulator, can be based on a SLG–insulator–SLG capacitor, which can switch one input signal from one output port to another. The main feature of the SLG–insulator–SLG capacitor switch is that it can enable or disable output ports by leveraging the charge accumulation in the capacitor by means of a voltage. This is an improvement over conventional current driven thermo- optic switches in Si photonics, which are operated under a continuous electrical current flow.

Thus, graphene photonics offers a combination of advantages in terms of both performance and manufacturing. The main material parameter to be optimized to achieve the best operating conditions is the carrier mobility. At mobilities >10,000 cm$^2$ V$^{-1}$ s$^{-1}$, a carrier concentration ~10$^{12}$ cm$^{-2}$ would ensure competitive modulation, detection and switching performances.

High mobility can be reached by using single crystals, an optimized transfer process and/or encapsulation. These aspects need to be combined with an optimized process to minimize the contact resistance. The remaining steps in the fabrication of graphene photonics coincide with those used for Si photonics integrated circuits. Given this, graphene photonics will offer a unique evolutionary pathway for photonics integration, with no technological discontinuity with respect to the existing and well-developed technologies.

We note that graphene is only one of thousands of possible layered materials[49]. In particular, transition metal dichalcogenides have a strong light–matter interaction and nonlinear optical effects. They could also be exploited for graphene encapsulation, instead of BN, to further increase graphene's mobility. These materials have been used as detectors[53] at non- telecom wavelengths and could also be used to make phase and electro- absorption modulators, as well as switches. However, their development is not yet at the stage of graphene-based devices.

**Appendix 1 | Basic concepts of optical modulation.**

The aim of a communication system is to transfer a message from one point to another[286]. Whether the message brings news to the receiver depends on the unpredictability of the message[286]. There is no point in transmitting a message if the receiver already knows its content. in digital communication systems messages are sent by modulating a source into sequences of bits[89]. The amplitude or phase of a lightsource can be used to encode the electrical signal into light that propagates along the optical channel. The most common method is binary encoding[287] by amplitude modulation, which is achieved by inducing ones (i.e. 'light on') and zeros (i.e. 'light off') by absorption or interference modulation[288]. The first case is known as electroabsorption modulation[61], and the second as Mach–Zehnder interferometer (MZI) modulation[61]. Phase modulation is an alternative used in complex modulation formats to achieve high-spectral-density[289] communication channels and maximize the ratio of data rate to spectral bandwidth.

In integrated photonics, the amplitude and phase can be modulated by acting on the electro-optical material that constitutes the waveguide[290] or, in the case of single- layergraphene (sLG), the material placed on top of the waveguide core[70–72,77,291]. The communication link is terminated with a receiver containing a photodetector. This system can discriminate an encoded signal, for example a binary signal, against the channel noise, and transfer the optical signal into a signal that can be processed by the electronics. The communication link is typically an optical fibre, and its performance is, among other factors, limited by the accumulated chromatic dispersion of the opticalfibre (ps $nm^{-1}$ $km^{-1}$ multiplied by the length of the link, which determines the intersymbol interference) and the power penalty[292] (the ratio of the average power required for a given value of extinction ratio to the power required for the ideal case of infiniteextinction ratio). The extinction ratio is the ratio of the signal power representing the logical bits '1' and '0' and is commonly expressed in dB. The average power is the mean of the power of the '1' and '0' bits. For example, if the power for the '1'-bit is 1 mW andthat of the '0'-bit is 0.5 mW, the extinction ratio is $10\log_{10}(2)$ ~3 dB, and the average power is 0.75 mW. A low extinction ratio indicates that a fraction of the power is un-modulated, which leads to a reduction in the receiver signal.

The term datacom describes communication within data centres, comprising links of short lengths (~2 km) according to the ethernet alliance. The term telecom is used forlonger links[85], from tens of kilometres to transoceanic distances. In datacom, link lengths are shorter, hence, smaller extinction ratios are tolerable in some cases, because the priority is to reduce size[293], insertion loss and power consumption[293].In telecoms, the penalty contributions arise from chromatic dispersion, channel losses, nonlinearities and accumulated amplified spontaneous emission noise of erbium doped optical amplifiers[85]. Although chromatic dispersion can be managed by a combination of appropriate signal coding and digital post-processing at the receiver[294] and losses can be compensated by optical amplifiers, nonlinearities and noise remaincrucial impairment factors[85].

**Appendix 2 | Modulators and detectors for optical transceivers**

State-of-the-art integrated modulators used in optical links are InP-based electro-absorption modulators[295–297], mainly for short-distance communications (up to ~80 km), Si-based photonic Mach–Zehnder interferometer (MZI) modulators[125] in pluggable modules (QSFP28) for <10 km-link-length interconnections (100GE CLR4, 100GE LR4 or 100GE ER4) or $LiNbO_3$-based MZ is for links >100 km. The parameters indicating modulator performance are energy consumption (pJ per bit), electrical bandwidth, insertion loss (IL) and extinction ratio (ER). IL is defined as the ratio between the optical power exiting the modulator when a '1'- bit is transmitted and that entering the modulator.

Another important parameter for modulators based on phase shifters (that is, those exploiting the modulation of the index of refraction[59,288]) is the voltage to induce a π phase shift, $V_\pi$, times the device length, L (that is, $V_\pi L$) given in dB cm. The figure of merit (FOM) of the phase shifting sections of MZI modulators includes the loss introduced by the component, $\alpha_{loss}$, given in dB $cm^{-1}$. For electro-refractive modulators, the FOM of the phase shifters is $FOM_{PM} = V_\pi L\alpha_{loss} = V_\pi \cdot IL$, expressed in VdB (ref.[298]), with IL= $L\alpha_{loss}$. In this case, a smaller $FOM_{PM}$ gives a higher performance. For electro-absorption modulators, $FOM_{EA} = ER/IL$[299]. the larger the $FOM_{EA}$, the better the performance.

Photodetectors have been developed for si photonics and for InP platforms. In both cases, they are integrated with an optical waveguide[287,288]. InP- based photodetectors are commercially available and widely adopted in most optical receivers[300]. Ge photodetectors are epitaxially grown on si waveguides[41] and can detect light from the waveguide through edge coupling if the Ge photodetectors terminate the si waveguide[301] or through evanescent coupling[41] if the Ge photodetectors are grown on the si waveguide[41]. The key performance indicators are responsivity, bandwidth and dark current. The responsivity, $R_{ph}$ (in A $W^{-1}$ or V $W^{-1}$), is defined as the electrical output of the device per optical input for the waveguide photodetector, excluding input coupling losses. The bandwidth is limited by the design of the circuit rather than by the material. The dark current is the leakage current that is present in absence of illumination. For Ge photodetectors[41] at an operating wavelength of 1,300 or 1,550 nm, a typical $R_{ph}$ is ~0.8 A $W^{-1}$ or 1 A $W^{-1}$, respectively.

Modulators and detectors are used in optical modules for transmitting and receiving signals. The optical interface specifications of three different 100GE (Gigabit ethernet standards — CLR4, LR4 (ref.[18]) and ER4 (ref.[18])) — are shown in the table below. CRL4, LR4 and ER4 indicate three interconnection lengths. the 100GE LR4, ER4 transceivers have a fibre ribbon leading four separate optical channels at a data rate of 25 Gb $s^{-1}$, at an operating wavelength range of 1,260–1,360 nm. 100GE CLRE CLR4 includes wavelength MUX and DEMUX (multiplexer and demultiplexer) and uses one single mode fibre to transmit and one to receive four wavelengths at 25 Gb $s^{-1}$.

Once the target optical link is chosen, the FOM can determine the specifications for the transmitter, such as the laser power or the driving voltage of the modulator. For example, consider a 100GE ER4 link used to perform a 100 Gb $s^{-1}$ transmission through a 40 km optical fibre (see the figure). The optical powers at the modulator input and output when '0' and '1' bits are transmitted are $P_{in}$, $P_{out,0}$ and $P_{out,1}$, and the power at the detector is $P_{detector}$. In the 100GE ER4 link, the receiver sensitivity, that is, the minimum detectable power, is $P_{detector}$ = −21.4 dBm , which corresponds to 0.0072 mW (with P(dB) = $10\log_{10}$[Pdetector (mW)/1 mW]). Considering a link loss of 16 dB, a dispersion penalty of 1 dB and a power penalty of 0.6 dB for ER = 8 dB (total 17.6 dB), the output power $_{Pout,1}$ = $P_{detector}$ + 17.6 dB = −3.8 dBm (ref.[281]). For ER= 8 dB, $P_{out,0}$ = $P_{out,1}$ − ER = −11.3 dBm. These power levels are then reflected in the transmitter specifications. If the transmitter is an electro-absorption modulator with $FOM_{EA}$ = ER/IL = 3 (ref.[111]), $P_{in}$ = $P_{out,1}$ + IL = −1.1 dBm. If a MZI with $FOM_{PM}$ = 12 v dB is used[125], $P_{in}$ can be found if the modulator peak- to-peak driving voltage, $V_{PP}$, is specified. If we assume $V_{PP}$ = 1 V, to be compatible with complementary metal oxide semiconductor (CMOs) technology[302], an ER = 8 dB implies V ~$0.37V_{\pi}$ (refs[133,134]) in push–pull configuration[303], with $V_{\pi}$ = 2.7 V ·IL =L $\alpha_{loss}$ = $FOM_{PM}/V_{\pi}$ = 4.4 dB, so that $P_{in}$ = $P_{out,1}$ +IL = 0.6 dBm.

| Application | Link length (km) | Link loss (dB) | Dispersion penalty (dB) | Extinction ratio (dB) | Power penalty (dB) | Receiver sensitivity (dB) |
|---|---|---|---|---|---|---|
| 100 GE CLR4 | 2 | <1 | <1 | 4-5 | 3 | -8.5 |
| 100 GE LR4 | 10 | <5 | Not applic. | 7 | 1.8 | -8.6 |
| 100 GE ER4 | 30-40 | 16 | ~1 | 8 | 0.6 | -21.4 |

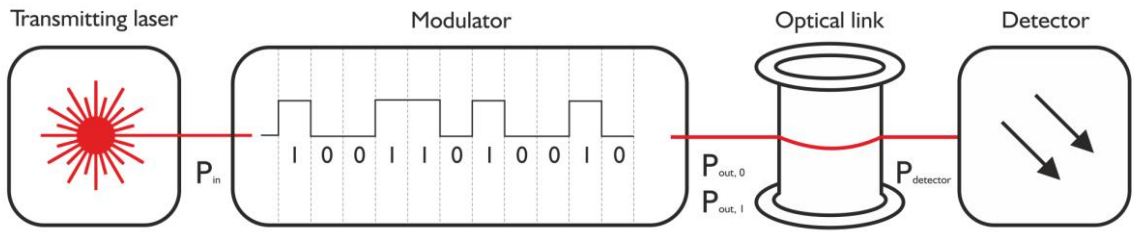

**Appendix 3 | the quality of graphene required for integrated photonic devices**

For graphene, the definition of 'quality' depends on the application. For example, the 'quality' needed for batteries and supercapacitors[304], has nothing to do with that required for electronic and optoelectronic applications[49]. As a result, it is important to clarify what this term means in the context of integrated photonic devices[49]. Such devices need the carrier mobility, μ, to be maximized[76]. Chemical vapour deposition (CVD) on Cu, substrate removal by $FeCl_3$, spinning of poly(methyl methacrylate) (PMMA) for transfer and subsequent PMMA removal are all steps that may release contaminants[269,305]. In addition, the transferred single-layer graphene (SLG) may have defects, wrinkles and/or non-uniformities. The interface between the SLG and the substrate may contain unpassivated dangling bonds that trap charges[271]. Each of these contributes a scattering time τ, which leads to a decay rate in the Kubo model[75]. The overall τ is the sum of individual contributions:

$$\tau^{-1} = (\tau_{phonons})^{-1} + (\tau_{short\text{-}range})^{-1} + (\tau_{charge\text{-}impurities})^{-1} + \ldots \qquad (6)$$

The shorter the τ, the more SLG absorbs light at large doping (Fig. 3). Eventually, this causes an increase in insertion loss and limits the maximum extinction ratio. The relation between τ and μ can be derived as follows: because τ does not depend on the frequency in the model of ref.[80], we consider ω → 0. In this limit, the intraband contribution to the complex conductivity of SLG becomes[70,80]:

$$\lim_{\omega \to 0} \sigma(\omega) = e^2 \frac{E_F}{\pi \hbar^2} \tau \quad (7)$$

where e is the electron charge, $k_B$ = 1.38 x $10^{-23}$ $m^2$ kg $s^{-2}$ $K^{-1}$ is Boltzmann's constant, T is the temperature, $n_s$ is the surface charge density and $E_F$ is the Fermi energy. Because $E_F = \hbar v_F \sqrt{\pi |n_S|}$ (ref.[70]), vF ≈ 9.5 x $10^7$ cm $s^{-1}$ and[306]

$$\lim_{\omega \to 0} \sigma(\omega) = n_S e \mu \quad (8)$$

From equations 7 and 8 we get

$$\mu \sim \frac{e v_F^2}{|E_F|} \tau \qquad \text{for } |E_F| \gg k_B T \quad (9)$$

From equation 9, τ ~10 fs, 100 fs and 300 fs correspond to μ ≈ 220, 2,200 and 6,600 $cm^2$ $V^{-1}$ $s^{-1}$ for $E_F$ = 0.4 eV, respectively. This range of values is typical for large-area SLG grown by CVD and wet transferred. Even if SLG fabrication and transfer are ideal, from equation 6, phonon scattering[282] sets an upper bound to τ = $\tau_{phonons}$, and consequently, the limit is $\mu_{phonons} = \mu(\tau_{phonons})$. Combining equations 8 and 9, we get:

$$\tau_{phonons} = \frac{\hbar^2 \pi}{e^2 E_F} \sigma_{phonons} \quad (10)$$

$$\mu_{phonons} = \frac{e v_F^2}{|E_F|} \tau_{phonons} \quad (11)$$

where the SLG conductivity can be assumed to be weakly dependent on $n_s$ around room temperature[307]. To a first approximation, $\sigma_{phonons}$ ≈ 1/50 Ω−1 (ref.[307]). At high doping ($n_s$ > 1 x $10^{13}$ $cm^{-2}$ and $|E_F|$ > 0.4 eV), equations 10 and 11 are lower estimates[282] and need experimental validation. We note that substrate engineering, and the encapsulation of SLG in different layered materials, may change the phonons involved in the scattering process, resulting in much higher mobilities than reported in Ref.[282] for a given doping.

**Appendix 4 | Mechanisms of optical modulation**

In a photonic circuit, optical modulation is obtained either by varying the absorption of the material through which propagation takes place or by varying its refractive index (n). The former case is known as electro-absorption modulation[59,143,308–311] and the latter as electro-refractive modulation. Phase modulation can be turned into amplitude modulation using a Mach–Zehnder interferometer (MZI)[312–316].

**Modulation in silicon**
*Plasma dispersion.* This occurs when Si absorbs a photon and an electron in the conduction band or a hole in the valence band is excited and occupies an available state in the same band[84]. This process may appear as absorption[108]. As a consequence of the Kramers–Kronig relations[317], both absorption and n vary with carrier concentration, N. In the case of Si, the following equations apply for variations in absorption (Δα) and refractive index (Δn) at 1.55 μm (ref.[108]

$$\Delta \alpha = 8.5 \times 10^{-18} \Delta N_e + 6.0 \times 10^{-18} \Delta N_h \quad (12)$$

$$\Delta n = -8.8 \times 10^{-22} \Delta N_e - 8.5 \times 10^{-18} (\Delta N_h)^{0.8} \quad (13)$$

*Franz–Keldysh effect.* In semiconductors, on application of an electrical field, the bands can be distorted, causing a shift in absorption[64,65,318], which can be used to modulate transmission[60,66]. In Si photonics, modulation through the Franz–Keldysh effect has been shown in GeSi alloys with <1% si, grown on si waveguides[60,66]. State-of-the-art GeSi electro-absorption modulators[319] integrated in Si photonics circuits operate at rates of up to 100 Gb $s^{-1}$.

*Quantum-confined Stark effect.* This is observed when an electrical field is applied to a quantum well[67]. In the absence of such a field, electrons and holes occupy a discrete spectrum of energy bands. the electric field modifies the bands, causing variations in absorption and n, analogous to the Franz–Keldysh effect[68].

**Modulation in graphene**

In graphene, α and n depend on $E_F$ and the intraband and interband transitions of electrons and holes excited by impinging photons[73,82,320,321]. In undoped SLG, the absorption of photons of any wavelength is allowed[69]. However, if $E_F$ is increased above half the photon energy, because of Pauli blocking[83], carrier excitation is inhibited, and SLG becomes transparent[83]. Electro-absorption modulation in SLG has been achieved by $E_F$ modulation[71,72]. $E_F$ modulation also causes phase modulation because α and n depend on $E_F$ (ref.[80]). When interband transitions are inhibited, absorption can occur only as a result of intraband transitions. These are primarily a consequence of long range scattering induced by, for example, impurities, trap states and screening. S convenient way of describing the overall effect of intraband transitions is the scattering time τ (see Appendix 3). The longer the τ, the lower the intraband absorption, that is, the more transparent SLG becomes in the $E_F$ range where interband transitions are excluded because of Pauli blocking. in this case, an $E_F$ modulation results in variations in n, thus enabling phase modulation[70,77,86].

**Appendix 5 | The evolution of mobile communications: the 5G network**

The evolution towards 5G will influence and define the amount and nature of data traffic in all network segments. Hence we discuss 5th generation wireless network concepts and not just component technologies aiming at high performance optical networks.

Fig.1a outlines the structure of a telecommunication network, highlighting the major segments: "Access", "Aggregation" with local-, metro- and regional networks, and "Core". Based on the traffic characteristics and topologies, these network segments have very different requirements in terms of cost, performance, quality and unit count for the devices which operate in them. This has consequences for their components and the underlying physical technologies. The data traffic is aggregated from the network edges (access) to the network core. The routers in the nodes of the core network need to process high bit rates ~several tens Tbit/s. At present, devices at the network periphery like, e.g., user equipment and residential home network terminations, usually need to deal with data rates of several tens to hundreds Mbit/s, and need to be affordable for the end-user and produced in billions of units. The predicted strong growth of data traffic over the coming years is mainly fuelled by the increasing demand for high definition (HD)-video applications, such as e.g. HD content streaming, -virtual reality, -augmented reality and - gaming. Today's users expect ubiquitous and anytime availability of these applications, which requires broadband wireless connections supporting extremely high bit rates to the user equipment (UE). 5G will address these demands for high bandwidth, together with the requirements coming from a different category of applications based on highly reliable low latency networking technologies described by the terms IoT or "Machine to Machine Communication, M2M"[324,325]. Typical applications are autonomous driving, industrial automation ("Industry 4.0"), smart-X (i.e. smart grid, smart home, …), e-health (Fig.1b). This class of applications is characterized by requirement of low transmission latencies down to 1 ms, high security and reliability, and low or moderate data rates (kbit/s) per device (sensor or actuator). However, the expected vast abundance of these devices (projected worldwide counts are hundreds of billions[324,325]) will give a significant contribution to the total traffic increase expected for the coming years. The 5G system will extend beyond previous generations of mobile communications, while ensuring a seamless upgrade for operator networks and quality of experience for the final users, with new services and enhanced network performance. To enable 5G, all the available network infrastructures should evolve with new levels of flexibility and automation (i.e. networks performing self-operations, optimization and healing), with higher priority given to network optimization, security, energy and cost efficiency[326,327]. 5G is designed to deliver, in a feature rich network, more diversified services with respect to previous generations.

Like its predecessors, the realization of the 5G network involves wireless as well as fixed link technologies (i.e. Cu-wire or fibre). Wireless links are present in the access part of the network, connecting radio antennas mounted on small- and large-cell stations with single connection, or interconnected peer to peer, end-users devices via radio waves or mm waves (wireless access links), point to point interconnections between high power cell towers and small cell stations, and amongst small cell stations (i.e. wireless fronthaul link).

There are two ways to have access to the 5G network: one is through a traditional access node, already defined in the previous generation of mobile communications, and the other is through a virtual access node. The latter is defined by the installation of a dedicated cell for industries or private applications, where wireless connectivity and mobility is performed with uniform industry-wide or private policy and mobility requirements. 5G through virtual access nodes can deliver more customized and dedicated network services with the quality and network performance settled by the final user.

Wireless access networks are composed of cell sites, divided into sectors, sending data through radio waves[328]. 4G long-term evolution (LTE) wireless technology is the foundation for 5G. Whereas LTE supports data rates up to

~150Mbit/s, LTE Advanced supports peak data rates up to ~1Gbit/s in the downlink. The practical individual user experience, limited by a number of technical constraints, usually is lower at several tens of Mbit/s for LTE. Unlike 4G, which requires mainly large, high-power, cell towers to radiate signals over longer distances (large towers in Fig. 1), 5G wireless access signals are for efficiency reasons transmitted also via a larger number of small cell stations located in places like light poles or building roofs. Cu-wires and/or optical fibres provide "fixed access" to buildings (like e.g. residential homes and industrial premises) or are part of the wireless infrastructure when connecting detached remote antenna heads to their base station controllers. Fixed links are present in the core and aggregation (metro) networks (defining the backhaul network, where telecom grade networks are connected directly with remote datacom or IP networks to have access to high computational resources), and close to the access networks (defining the wired fronthaul network, where telecom grade networks are directly connected with datacom networks, closer to the final user, so to be fast and responsive in delivering latency-sensitive services to the final user) (Fig. 1a). Typically, optical fibres ensure broadband connectivity from the core up to radio antennas and end-user premises (e.g. house, office, buildings, campus and stadiums).

Two main scenarios are seen in the 5G network evolution. 1) Mobile wireless cloud networks leading to the concepts of Smart City, Smart Industry, and Smart Life applications. These are evolving and need to ensure high mobility and network scalability, while supplying enhanced connectivity. 2) Dedicated networks for end-user premises applications. These leverage both broadband connectivity based on optical interconnections to the rest of the network and pervasive wireless networks (e.g. WiFi; Visible Light Communication Systems, such as Light Fidelity (LiFi); Mobile Indoor Localization Systems).

In the mobile wireless cloud network the worldwide available connections can transmit on average 100Mbit/s per user. The download of one HD movie (e.g. 1500Mb) can take less than two minutes. The dedicated networks to the final user premises enable the worldwide access to a 5G network with an average capacity of 1Gbit/s per user, thus, the download of an HD movie can take ~10 seconds.

5G networks will give people access to a networked society, where anybody will want to communicate with more people, in more contexts and for more reasons than ever before. In the networked society, communications will matter even more than in the past, since the global economy will increasingly depend on our ability to communicate. 5G enables the telecom and datacom networks to provide new services to the benefit of operators and their customers. A huge number of new communication services will be created to satisfy the increasing demand for new ways to communicate. As a consequence, new communication services will be continuously proposed and new network models will be assessed.

To track the 5G shift, several trends illustrating the impact on communication services are represented in Fig 1.
1) Strong growth in the number of connections through subscription in emerging markets, with an increase in the number of devices per person (13Bn by 2021: smartphones, laptops, tablets), as well as an increased number of machines and equipment connected, with the rising IoT (number of IoT devices expected to grow by 15Bn by 2021)[329]. A large numberand different types of objects with IP address will be controlled, monitored, connected through the network. Thus, the IoT aggregate data traffic will overcome that of traditional voice and data communications.
2) Smarter and more capable devices, including smartphones and tablets, as well as Smart TVs. Each has the potential to radically increase our ability to communicate in different ways. Smartphones are the new global standard. Tablets also serve as communication hubs in the home. Smart TVs opened new forms of communication in a previously unexplored context.
3) Consumer demands for communication continuously increasing, as people come to expect high quality and coverage for less money. They are also communicating more than ever before, but not always using traditional voice or SMS channels. As the internet is becoming communication-centric, communication channels have started to blur, and consumers will be more open to migrating traditional phone-related behaviour to other technological platforms.
4) Expansion of communications into ever more contexts through social media interactions, ubiquitous video and, increasingly, machine-to-machine with the IoT rise.

With 5G there is an opportunity for telecom and datacom operators to put communications in different contexts with new business models. They can reach out to users not in their customer base (through service exposure, context-based and machine-to-anything communications). E.g., within the automotive sector, if a car crashes, the release of the airbag could trigger a direct connection to the emergency services, as well as to a towing service and garage. The user would pay for the service, with the operator acting as frontend or interface. There could be similar situations with communications included in home alarms, gaming, transport, mobility and in-web communications, smart trash collection and reuse.

Today people are better informed, have more choices and are in a better position to be demanding to their service providers. This puts greater pressure on operators to offer consumers more value for money. At the same time, the phone is becoming a multicommunications device that includes all communications trends.

Smartphones have made the use of complex services controllable for everyone. Besides the price, flexibility and accessibility define the service quality perception. Operators' networks need to evolve on performance and flexibility to support this reality of rich, varied, and often free, services. A leading example is the evolution of domotics and home security, influencing everyday life. 5G networks could enable automation and reliable remote control in real time of the main basic home functionalities, from the optimization of energy and resource utilization, to lowering the impact of domestic waste by performing optimized recycling processes through automated waste selection and proper packing.

The 5G network also supports interactive and portable media players (e.g. mp3 reader, iPod), mobile equipment with sensors (e.g. environment monitoring or body temperature and blood pressure sensors), mobile wireless cameras to take pictures or video with high definition or for surveillance purposes, all devices capable of offering HD-video streaming, webmail delivery and any kind of internet connection. With 5G, specifically tailored services can be delivered, such as those dedicated to automotive, autonomous vehicles, including cars or buses, and apps for entertainment, traffic monitoring, chartering on demand of usual and unusual paths. To support users with time- and energy-optimized travel management, internet connectivity will be available even for users travelling with fast moving vehicles for public and private transportation, like airplanes or trains, thus redefining the concept of transportation as functional smart transportation (Fig.1b).

The evolution of 5G is creating ecosystems bringing different partners together offering different services (e.g. infrastructure, transport and building services) leveraging on the combined use of different technologies. All current and future companies can be influenced by the diffusion of capillary telecom and datacom networks, giving rise to an interconnected industrial evolution. Innovation and related business can be generated by ecosystems where small, medium and large enterprise interests proactively converge on 5G.

Finally, we consider the example of smart grids, where renewable and not renewable energy production, transmission, transformation, distribution and delivery are performed in networked infrastructures. E.g., environmental energy harvesting (clean energy from natural sources, such as sun and wind) and delivery needs to be optimized, adapting the network infrastructure performance to the real energy availability and final user demand. 5G IoT solutions introduce full control, manageability and surveillance (e.g. through sensors and smart meters, dedicated data centres for data storage and real time analytics) at sustainable cost and energy efficiency for operability of each element in the network (solar cells and energy collectors, electricity pylons, power lines equipped with sensors and transceivers for system monitoring), to the advantage of all industries, from primary electric suppliers over the national grid, to medium and small utility companies for local and regional infrastructures. The smart grid enables telecom operators to install 5G networks powered by renewable and clean energy sources, reducing the telecom and datacom network environmental footprint.

This industrial evolution towards smart manufacturing is enabled by robotics, machine intelligence and 5G, to improve productivity, while ensuring sustainable growth margins and speeding the delivery of new services for industry and people. The factory is going to be redefined through the digitization of the manufacturing process and plants, enabled by 5G networks, thus realizing new industrial scenarios where real-time communication between humans, robots, remote controlled machinery, factory logistics, and products are key[326].

**Related Links**

CLR4: https://www.clr4-alliance.org/

Ethernet Alliance: http://www.ethernetalliance.org/

Information and Communication Technologies, Environmental Sustainability and Climate Change: http://www.itu.int/en/action/climate/Pages/default.aspx

International Technology Roadmap for Semiconductors 2.0, 2013 Edition, Design: http://www.itrs2.net/2013-itrs.html

International Telecommunications Union: Interfaces for the optical transport network: https://www.itu.int/rec/ T- REC-G.709/en

International Telecommunications Union: Support of IP-based services using IP transfer capabilities: https://www.itu.int/rec/T- REC-Y.1241/en

Internet of Everything: https://newsroom.cisco.com/ioe

New 2018 Ethernet roadmap looks to future speeds of 1.6 terabits/s. inside HPC: https://insidehpc.com/2018/03/new-2018-ethernet-roadmap-looks-future-speeds-1-6-terabits-s/

Photonics Component & Circuit Design Software: https://www.lumerical.com/tcad- products/

Si- on-insulator (SOI): https://order.universitywafer.com/default.aspx?cat=Silicon%20on%20Insulator%20(SOI)%20wafers

The ethernet roadmap: https://newsroom.cisco.com/ioe

Xilinx: https://www.xilinx.com/publications/about/3-D_Architectures.pdf